\documentclass{article}

\usepackage{setspace}
\usepackage{fullpage}
\usepackage{graphicx}
\usepackage{amsmath}
\usepackage{hyperref}
\usepackage{authblk}
\usepackage{pdfpages}
\usepackage{txfonts}
\usepackage[nocompress]{cite}
\usepackage[font=small,labelfont=bf]{caption}

\makeatletter \renewcommand\@biblabel[1]{#1.} \makeatother

\begin{document}

\title{Information flow reveals prediction limits in online social activity}

\author[1,2,*]{James P.~Bagrow}
\author[1,2]{Xipei Liu}
\author[1,2,3,$\dagger$]{Lewis Mitchell}
 \affil[1]{\normalsize Department of Mathematics \& Statistics, University of Vermont, Burlington, VT, United States}
 \affil[2]{Vermont Complex Systems Center, University of Vermont, Burlington, VT, United States}
 \affil[3]{School of Mathematical Sciences, North Terrace Campus, The University of Adelaide, SA 5005, Australia}
 \affil[*]{E-mail: james.bagrow@uvm.edu}
 \affil[$\dagger$]{E-mail: lewis.mitchell@adelaide.edu.au}

\date{January 21, 2019}

\maketitle

\doublespacing

\begin{abstract}
Modern society depends on the flow of information over online social networks, 
and users of popular platforms generate significant behavioral data about themselves and their social ties\cite{Kossinets2006,lazer2009computational,Kwak2010,Bakshy2015,Garciae1701172}.
However,
it remains unclear what fundamental limits exist when using these data to predict the activities and interests of individuals, and to what accuracy such predictions can be made using an individual's social ties.
Here we show that 95\% of the potential predictive accuracy for an individual is achievable using their social ties only, without requiring that individual's data.
We use information theoretic tools to estimate the predictive information within the writings of Twitter users,
providing an upper bound on the available predictive information that holds for any predictive or machine learning methods.
As few as 8-9 of an individual's contacts are sufficient to obtain predictability comparable to that of the individual alone.
Distinct temporal and social effects are visible by measuring information flow along social ties, allowing us to better study the dynamics of online activity.
Our results have distinct privacy implications: information is so strongly embedded in a social network that in principle one can profile an individual from their available social ties even when the individual forgoes the platform completely.
\end{abstract}

\bigskip

The flow of information in online social platforms is now a significant factor in protest movements, national elections, and rumor and misinformation campaigns\cite{shirky2011political,lotan2011arab,delvicario2016}.
The study of social contagion\cite{Castellano2009}, for example, is predicated on the flow of  information over social ties,
and has benefited greatly from the availability of massive online social datasets and platforms on which to perform observational and experimental studies\cite{Kramer2014,Monsted2017}.
Data collected from online social platforms are a boon for researchers\cite{lazer2009computational} but also a source of concern for privacy, as the social flow of predictive information can reveal details on both users and non-users of the platform\cite{Garciae1701172,Jurgens2017,Garcia2018}.
Measuring information flow is challenging, in part due to the complexity of natural language and in part due to the difficulty in defining a quantitative and objective measure of information.
Owing to these challenges, proxies are often studied instead:
\emph{structural} proxies focus on network characteristics such as the movements of keywords\cite{lotan2011arab,gruhl2004information,bakshy2012role,Bakshy2015} or adoptions of behaviors\cite{aral2009distinguishing,centola2010spread,aral2012identifying}.
\emph{Temporal} proxies attempt to quantify the information contained in the timings of user activity, as temporal relationships between user activity are known to reflect underlying coordination patterns\cite{ver2012information,borge2016dynamics}. 

However, neither of the above approaches consider the full extent of information available: 
both the complete language data provided by individuals and their temporal activity patterns.
Although, for example, temporal proxies are necessary in social networks where time series data are available but message content is not, for privacy or other reasons (for example, in mobile phone datasets),
public postings to online social platforms present a unique opportunity to explore the textual content of messages in conjunction with their timings, giving a richer understanding of social ties.

Information theory allows us to mathematically quantify the information contained within data, and is well suited to data in the form of online written communication.
Although the mathematical definition of information is somewhat distinct from our commonly held notions of information and meaning, or semantics, information-theoretic measures are crucial for understanding how algorithms can learn from data.
Nowadays, with such large volumes of data generated by online social platforms, both researchers and platform providers are often forced to interact with a platform's data only computationally, using algorithms to quantify and make inferences about users, and the accuracy of these inferences is predicated on the mathematical information contained within a user's data.

In this work, we apply information-theoretic estimators to study information and information flow within a collection of Twitter user activities. 
These estimators fully incorporate language data while also accounting for the temporal ordering of user activities.
We find that meaningful predictive information about individuals is encoded in their social ties,
allowing us to determine fundamental limits of social predictability, independent of actual predictive or machine learning methods.
We explore the roles of information recency and social activity patterns, as well as structural network properties such as information homophily between individuals.

We gathered a dataset of $N = 13,905$ users, comprising egocentric networks from the Twitter social media platform,
and a total of $m = 30,852,700$ public postings from these users.
Each of the $n = 927$ ego-networks consisted of one user (the ego) and their $15$ most frequently mentioned Twitter contacts (the alters), providing us with ego-alter pairs on which to measure information flow.
See `Data collection and filtering' in the Methods section for full details on the data processing.

The ability of a machine learning method to accurately profile individuals from their online traces is reflected in the predictability of their written text.
Indeed, with a language model trained to predict the words a user will post online, in principle, one can construct a profile of the user by evaluating the likelihoods of various words to be posted, such as terms related to politics.
Thus, quantifying the predictive information contained within a user's text allows us to understand the potential accuracy such methods can potentially achieve given a user's data.

\begin{figure}[t]
\includegraphics[width=\textwidth]{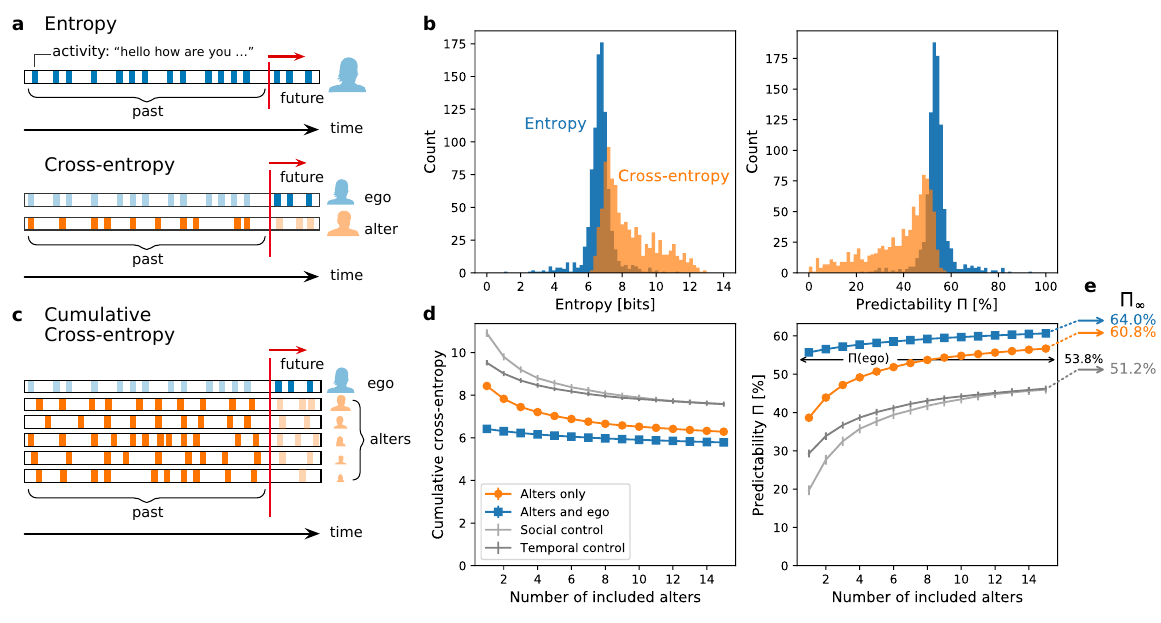}
\caption{\textbf{Information and predictability in online social activity.}
\textbf{a},
A user posts written text over time and we would predict their subsequent words given their past writing.
Treating each user's posts as a contiguous text stream, the entropy rate tells us how uncertain we are about a user's future writing given their past. 
To study information flow, the cross-entropy rate tells us how much information about the future text of one user (the ego, blue) is present in the past text of another user (the alter, orange).
\textbf{b},
Most users have entropies and predictabilities (blue) in a well-defined range, whereas the cross-entropies and associated predictabilities (orange) indicate a broad variety of social information flow levels.
\textbf{c},
Predictive information may be available in the pasts of multiple alters, so we computed the cumulative cross-entropy as we included more alters in order of most to least frequently contacted.
\textbf{d},
As the past activities of more alters are used to predict the ego, more information is available and the entropy drops and predictability rises (orange). 
Including the ego's past with the alters (blue) shows that the alters provided non-redundant predictive information.
\textbf{e},
Extrapolating beyond our data window estimates the prediction limit $\Pi_\infty$ of online activity.
Error bars denote mean $\pm$ 95\% CI.
\label{fig:introFig}
}
\end{figure}

A text's predictive information can be characterized by three related quantities, the entropy rate $h$, the perplexity $2^{h}$, and the predictability $\Pi$.
The entropy rate quantifies the average uncertainty one has about future words given the text one has already observed (Fig.~\ref{fig:introFig}a). 
Higher entropies correspond to less predictable text and reflect individuals whose interests are more difficult to predict.
In the context of language models, it is also common to consider the perplexity.  
Whereas the entropy rate specifies how many bits $h$ are needed on average to express subsequent  unseen words given the preceding text, the perplexity tells us that our remaining uncertainty about those unseen words is equivalent to that of choosing uniformly at random from among $2^{h}$ possibilities. 
For example, if $h=6$ bits (typical of individuals in our dataset), the perplexity is 64 words, which is a significant reduction from choosing randomly over the entire vocabulary (social media users have ${\approx}5000$-word vocabularies on average; see Supplementary Note 1.3 for full distributions). 
Finally, the predictability $\Pi$, given via Fano's inequality\cite{CoverThomas}, is the probability that an \emph{ideal} predictive algorithm will correctly predict the subsequent word given the preceding text.
Repeated, accurate predictions of future words indicate that the available information can be used to build profiles and predictive models of a user's writing (see also below for subsequent discussion), and estimating $\Pi$ allows us to fundamentally bound the usefulness of the information present in a user's writing without depending on the results of specific predictive algorithms.
We emphasize that the information-theoretic \emph{predictability} as defined here is distinct from \emph{prediction}, in that it does not actually make predictions about future text. 
Instead, this predictability provides a method-independent upper bound on prediction accuracy.

Information theory has a long history of estimating the mathematical information content of text\cite{shannon1951prediction,brown1992estimate,schurmann1996entropy,kontoyiannis_nonparametric_1998}.
Crucially, information is present not just in the words of the text but also in their order of appearance. 
Thus, we applied a nonparametric entropy estimator that incorporates the full sequence structure of the text\cite{kontoyiannis_nonparametric_1998}. 
This estimator has been proved to converge asymptotically to the true entropy rate for stationary processes and has been applied to human mobility  data\cite{song2010limits}. 
See `Measuring the flow of predictive information' and `Estimator convergence on our data' in the Methods section for further details on the entropy estimators and their convergence rates on these data.

We focus on four aspects of information flow over social networks, exploring both content and timing of messages: 
(i) the extent to which information is encoded through language into an individual's social ties,
(ii) the importance of recency to information flow between individuals,
(iii) the role of tie strength between individuals in the flow of information, and
(iv) the relationship between structural network properties such as homophily and information.
We first examined the information content of the egos themselves.
Their text streams were relatively well clustered around $h\approx 6.6$ bits, with most falling between 5.5--8 bits (Fig.~\ref{fig:introFig}b). 
Equivalently, this corresponds to a perplexity range of ${\approx}45$--$256$ words, far smaller than the typical user's ${\approx}5000$-word vocabulary, and a mean predictability of ${\approx}53\%$, quite high for predicting a given word out of ${\approx}5000$ possible words on average (for example, choosing words uniformly at random corresponds to a predictability of $0.02\%$).
We found this typical value of information comparable to other sources of written text, but social media texts were more broadly distributed around the mean---individuals were more likely to be either highly predictable or highly unpredictable compared with formally written text
(see Supplementary Note 1.4).

Next, instead of using the entropy rate to ask how much information is present in what the ego has previously written regarding what the ego will write in the future, we now ask how much information is present on average in what the \textbf{\textit{alter}} has previously written regarding what the ego will write in the future (Fig.~\ref{fig:introFig}a). 
If there is consistent, predictive information in the alter's past about the ego's future, especially beyond the information available in the ego's own past, then there is evidence of predictive information flow.

Replacing the ego's past writing with the alter's past converts the entropy to the \emph{cross-entropy} (see `Measuring the flow of predictive information' and `Estimator convergence on our data' in the Methods section). 
The cross-entropy is always greater than the entropy when the alter provides less information about the ego than the ego, and so an increase in cross-entropy tells us how much information we lose by only having access to the alter's information instead of the ego's.
Indeed, estimating the cross-entropy between each ego and their most frequently contacted alter (Fig.~\ref{fig:introFig}b), we saw higher cross-entropies than using the ego's own text, spanning from 6--12 bits compared with 5--9 bits
(equivalently, perplexities from 64--4096 words compared with 32--512 words, or predictabilities spread from 0--60\% compared with 40--70\%).
Whereas less frequently contacted alters provided less predictive information than alters in close contact (see Supplementary Notes 1.6 and 1.7), even for the closest alters
there was a broader range of cross-entropies than the entropies of the egos themselves.
This implies a diversity of social relationships: sometimes the ego is well informed by the alter, leading to a cross-entropy closer to the ego's entropy, whereas other times the ego and alter exhibit little information flow.

Thus far we have examined the information flow between the ego and individual alters, but actionable information regarding the future of the ego may be embedded in the combined pasts of multiple alters  (Fig.~\ref{fig:introFig}c).
To address this, we generalized the cross-entropy estimator to multiple text streams (see `Measuring the flow of predictive information' and `Estimator convergence on our data' in the Methods section).
We then computed the cross-entropies and predictabilities as we successively accumulated alters in order of decreasing contact volume (Fig.~\ref{fig:introFig}d).
As more alters were considered, cross-entropy decreased and predictability increased
(Spearman's $\rho = -0.505$ 95\% CI [-0.517, -0.492], $p < 0.001$ and $\rho = 0.527$ [0.515, 0.540], $p < 0.001$, respectively),
which is sensible as more potential information is available. 
Interestingly, with  8--9 alters, we observed a predictability of the ego given the alters at or above the original predictability of the ego alone---with 10 alters, the predictability was significantly greater than that of the ego alone (two-tailed test, $t(1852) = -3.32$, $p < 0.001$).
As more alters were added, up to our data limit of 15 alters, this increase continued. 
Paradoxically, this indicated that there is potentially more information about the ego within the total set of alters than within the ego itself.

To understand this apparent paradox, we need to address a limitation with the above analysis: it does not incorporate the ego's own past information. 
It may be that the information provided by the alters is simply redundant when compared to that of the ego.
To see whether this is the case, we simply included the ego's past alongside the alters, generalizing the 
estimator to an entropy akin to a transfer entropy\cite{schreiber2000measuring,staniek2008symbolic}, a common approach to studying information flow.
This entropy is computed in the `Alters and ego' curves in Fig.~\ref{fig:introFig}d. 
A single alter provided a small amount of extra information beyond that of the ego, $1.9\%$ more predictability. 
This value provided us a quantitative measure of the extent of information flow between individual users of social media.
Beyond the most frequently contacted alter, as more alters were added, this extra predictability grew: at 15 alters and the ego there was $6.9\%$ more predictability than via the ego alone. 
Furthermore, the information provided by the alters without the ego is strictly less than the information provided by the ego and alters together, resolving the apparent paradox.

However, this extra predictability also appeared to saturate, and if so then eventually adding more alters will not provide extra information (see Supplementary Note 1.2). 
This observation is compatible with \emph{Dunbar's number} which uses cognitive limits to argue for an upper bound on the number of meaningful ties that an ego can maintain (${\approx}150$ alters for humans)\cite{dunbar1993coevolution}.
Thus, the question becomes: given enough ties, what is the upper bound for predictability?

To extrapolate beyond our data window, we fitted a nonlinear saturating function
to the curves in Fig.~\ref{fig:introFig}d,  
(see Supplementary Note 1.2 for details and validation of our extrapolation procedure). 
From fits to the raw data extrapolated to infinity, we found a limiting predictability given the alters of 
$\Pi_{\infty} =  60.8\% \pm 0.691$\% (95\% CIs) (Fig.~\ref{fig:introFig}e). 
Of course, egos will not have an infinite number of alters, so a more plausible extrapolation point may be to Dunbar's number:  
$\Pi_{150} = 60.3$\%, 
within the margin of error for $\Pi_{\infty}$, indicating that saturation of predictive information has been reached.
Similarly, extrapolating the predictability including the ego's past gives 
$\Pi_{\infty} = 64.0\% \pm 1.54\%$ 
($\Pi_{150} = 63.5\%$).

These extrapolations showed that significant predictive information was available in the combined social ties of individual users of social media. 
In fact, there is so much social information that an entity with access to all social media data will have only slightly more potential predictive accuracy (${\approx}64\%$ in our case) than an entity that has access to the activities of an ego's alters but not to those of that ego (${\approx}61\%$). 
This may have distinct implications for privacy: if an individual forgoes using a social media platform or deletes their account, yet their social ties remain, then, potentially, that platform owner still possesses $95.1\% \pm 3.36\%$ of the achievable predictive accuracy of the future activities of that individual.

Two issues can affect the cross-entropy as a measure of information flow. 
The first is that the predictive information may be due simply to the structure of English: commonly repeated words and phrases will represent a portion of the information flow. 
The second is that of a common cause: egos and alters may be independently discussing the same concepts. 
This is particularly important on social media with its emphasis on current events\cite{Kwak2010}.

To study these issues, we constructed two types of controls. 
The first randomly pairs users together by shuffling alters between egos. 
The second constructed pseudo-alters by assembling, for each real alter, a random set of posts made at approximately the same times as the real alter's posts, thus controlling for temporal confounds. 
See `Control procedures' in the Methods section for more information.
Both controls used real posted text and only varied the sources of the text.
As shown in Fig.~\ref{fig:introFig}d, the real alters provided more social information than either control. 
Although there was a decrease in entropy as more control alters were added, the control cross-entropy remained above the real cross-entropy 
(two-tailed test, $t(23293) = -103.8$, $p < 0.001$) 
and the control predictability remained below the real predictability
($t(21103) = 119.0$, $p < 0.001$).
We also observed that, for a single alter, the temporal control had a lower cross-entropy than the social control 
($t(23293) = -117.5$, $p < 0.001$) 
and therefore temporal effects explain more information than social effects (underscoring the role of social media as a news platform\cite{Kwak2010}), although both controls eventually converge to a limiting predictability of $51.2\%$.
This demonstrates that useful predictive information is encoded in real social ties, beyond that expected from the structure of language alone.

\begin{figure*}[t]
\centering
\includegraphics[width=0.5\textwidth]{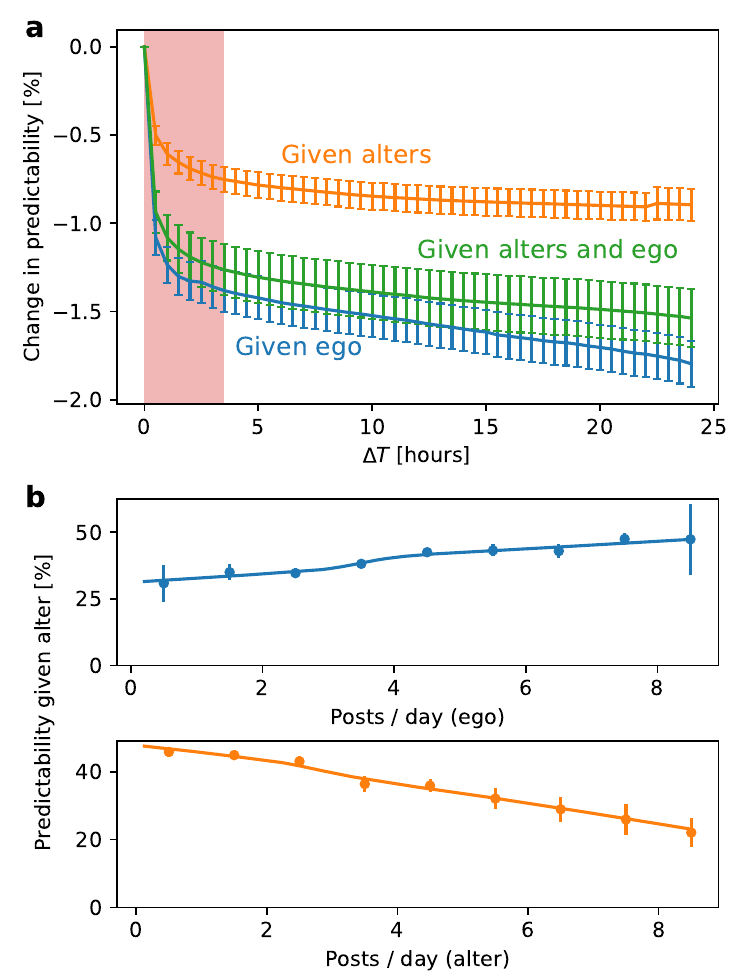}
\caption{\textbf{Recency of information.} 
\textbf{a},
Removing the most recent $\Delta T$ hours of activity, most predictive information about the ego is contained in the most recent 3--4 h (shaded region, $1.4\%$ drop).
In all cases, information extends backwards beyond these time intervals, but the ego (blue) contains more long-range past information ($0.6\%$ more predictability) than the combined alters alone (orange, alters and ego green).
\textbf{b},
Egos who post more frequently are $17\% \pm 14.9\%$ more predictable from their alter than egos who post less frequently, whereas frequently posting alters provide $23\% \pm 4.46\%$ less information about their egos than alters who post less often.
Lines in panel b denote a LOWESS fit.
Error bars denote mean $\pm$ 95\% CI.
\label{fig:recency}
}
\end{figure*}

Given the importance of temporal information in online activity, 
to what extent is this reflected in the information flow?
Do recent activities contain most of the predictive information or are there long-term sources of information? 
To estimate recency effects, we applied a censoring filter to the ego's text stream, removing at each time period the text written in the previous $\Delta T$ hours and measuring how much the mean predictability decreased compared with the mean predictability including the recent text.
Increasing $\Delta T$ decreased $\Pi$, especially evident when removing the first 3--4 h worth of text (these intervals correspond to 6.2--7.8 tweets ignored per word on average; Fig.~\ref{fig:recency}a): we found an average decrease in predictability of $1.4\%$ at 4 h. 
This 1.4\% loss in predictability relative to the uncensored baseline is comparable to the $1.9\%$ gain from the rank-1 alter that we observed in Fig.~\ref{fig:introFig}d.
In other words, close alters tended to contain a quantity of information about the ego comparable to the information within just a few hours of the ego's own recent past.
Beyond 24 h the predictability loss continued approximately linearly (visually; see Supplementary Note 1.5 and Supplementary Fig.~8).
We also applied this censoring procedure to the alters alone and the alters combined with the ego, excluding their recent text and measuring how the cross predictability changed on average from their respective baselines.
We found a similar drop in predictability during the first few hours ($0.8\%$ and $1.3\%$ in the first 4 h given alters and alters plus ego, respectively), but then a more level trend than when censoring the ego alone (a further decrease of $0.1\%$ and $0.3\%$, respectively, between 4 and 24 h, compared with $0.4\%$ for the ego alone over the same interval). 
This leveling off showed that less long-term information was present in the alters' pasts than within the ego's.

Next, we studied recency by the activity frequencies of alters and egos.
Individuals who post frequently to social media, keeping up on current events, may provide more predictive information about either themselves or their social ties than other, infrequent posters.
We found that the self-predictability of users was actually independent of activity frequency (Supplementary Note 1.4), but there were strong associations between activity frequency and social information flow:
egos who posted 8 times per day on average were $17\% \pm 14.9\%$ (95\% CI) more predictable given their alters than egos who posted once per day on average (Fig.~\ref{fig:recency}b).
Interestingly, this trend reversed itself when considering the activity frequencies of the alters:
alters who posted 8 times per day on average were $23\% \pm 4.46\%$ less predictive of their egos than alters who posted once per day on average. 
Both trends in Fig.~\ref{fig:recency}b were significant (Spearman's $\rho = 0.276$ [0.216, 0.335], and $\rho = -0.437$ [-0.487, -0.383], respectively, $p < 0.001$; see Supplementary Note 1.6).
Highly active alters tended to inhibit information flow, perhaps due to covering too many topics of low relevance to the ego.

\begin{figure*}[t]
\centering
{\includegraphics[width=0.5\textwidth]{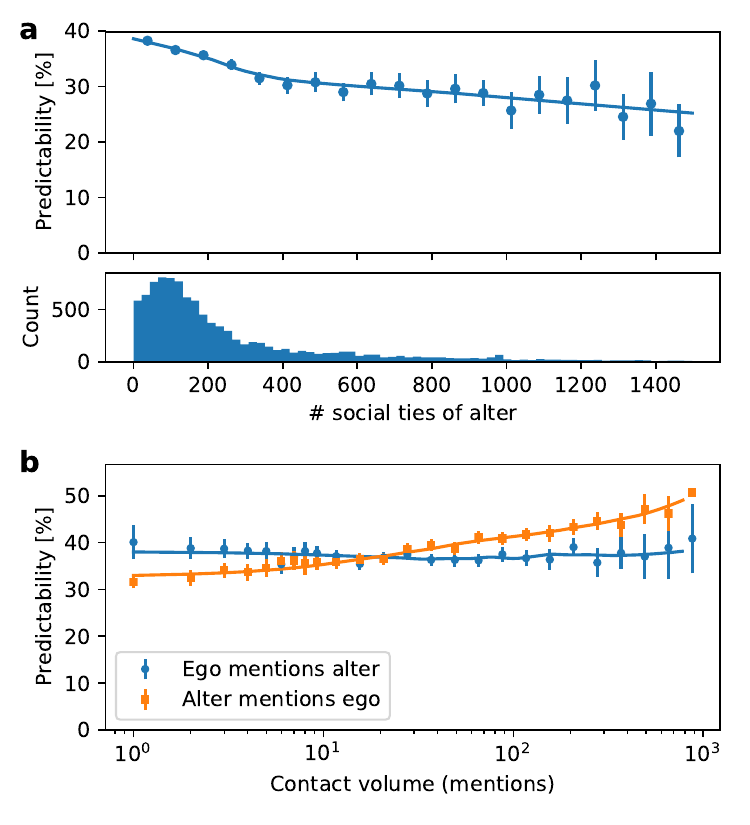}}
\caption{\textbf{Social interactions are visible in information flow.}
\textbf{a},
Alters with more social ties of their own provided less information about the ego than less popular alters (Spearman's $\rho = -0.199$ [-0.224, -0.175], $p < 0.001$).
\textbf{b},
Information flow captures directionality in relationships, which is a key factor in social dynamics.
Alters who often contact the ego provide more predictive information about the ego than alters who rarely mention the ego (Spearman's $\rho = 0.226$ [0.202, 0.250], $p < 0.001$).
Yet, if the ego frequently mentions the alter, it does not necessarily mean that the alter will provide more predictive information about the ego (Spearman's $\rho = -0.0185$ [-0.0440, 0.00704], $p=0.156$).
Lines denote a LOWESS fit.
Error bars denote mean $\pm$ 95\% CI.
\label{fig:mentionsVsKLeaAndKLae}
}
\end{figure*}

Information flow reflects the social network and social interaction patterns (Fig.~\ref{fig:mentionsVsKLeaAndKLae}).
We measured information flow for egos with more popular alters compared with egos with less popular alters.
Alters with more social ties provided less predictive information about their egos than alters with fewer ties (Fig.~\ref{fig:mentionsVsKLeaAndKLae}a).
This trend was significant (Spearman's $\rho = -0.199$ [-0.224, -0.175], $p < 0.001$; see Supplementary Note 1.9).
Qualitatively, the decrease in predictability of the ego was especially strong up to alters with ${\sim}400$ ties, where the bulk of our data lies, but the trend continued beyond this as well.
This decreasing trend belies the power of hubs in many ways: although hubs strongly connect a social network topologically\cite{Reka2000}, limited time and divided attention across their many social ties bound the hub alter's ability to participate in information dynamics mediated by the social network and this is reflected in the predictability.

Reciprocated contact is an important indicator of social relationships\cite{wasserman1994social}, 
especially in online social activity where so much communication is potentially one-sided\cite{Kwak2010}.
In Fig.~\ref{fig:mentionsVsKLeaAndKLae}b, we investigated how directionality in contact volume, how often the ego mentions the alter and vice versa, related to information flow. 
We found that the ego was more predictable given the alter for those dyads in which the alter more frequently contacted the ego (Spearman's $\rho = 0.226$ [0.202, 0.250], $p < 0.001$; see Supplementary Note 1.9), but there was little change across dyads when the ego mentioned the alter more or less frequently (Spearman's $\rho = -0.0185$ [-0.0440, 0.00704], $p = 0.156$; see Supplementary Note 1.9). 
We also observed a similar trend for information flow but in reverse, when predicting the alter given the ego (see Supplementary Note 1.9.
These trends captured the reciprocity of information flow: an alter frequently contacting an ego will tend to give predictive information about the ego, but the converse is not true: an ego can frequently contact her alter but that does not necessarily mean that the alter will be any more predictive, as evidenced by the relatively flat trend in Fig.~\ref{fig:mentionsVsKLeaAndKLae}b.

\begin{figure}[t]
\centering
\includegraphics[scale=0.65]{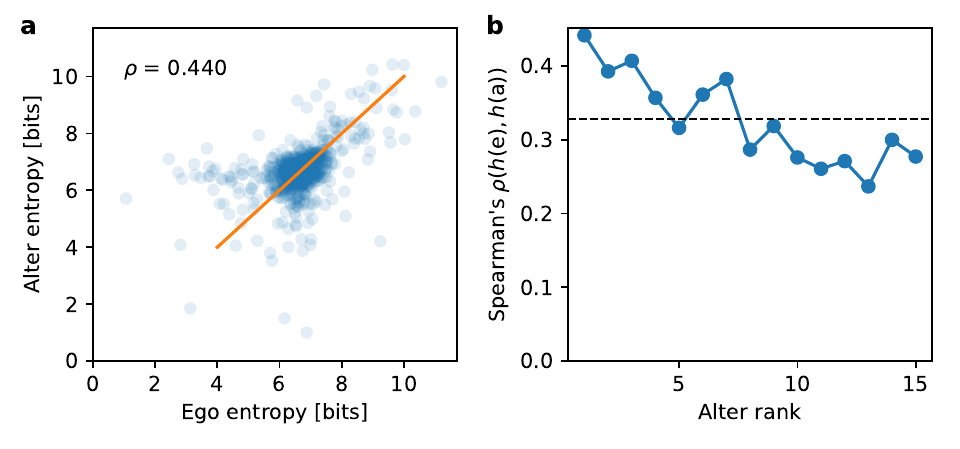}
\caption{\textbf{An `information homophily' between egos and alters.}
The entropies of egos and their alters are strongly correlated (Spearman's $\rho = 0.440$ [0.386, 0.490, $p < 0.001$), indicating a \textbf{\textit{homophily effect}}.
\textbf{a}, The entropy rate $h$ of egos compared to the entropy rate of their rank-1 (most frequently contacted) alter. 
The straight line $y=x$ provides a guide for the eye.
\textbf{b}, The Spearman's $\rho$ between ego entropies and alter entropies as a function of alter rank; all plotted $\rho$ are significant ($p < 0.001$). 
Correlation generally decreases with alter rank. 
The dashed line indicates $\rho$ across all ranks.
\label{fig:infoHomophily}
}
\end{figure}

Finally, comparing the entropy of an ego with the entropy of their alters reveals a strong homophily effect in terms of their (self) information (Fig.~\ref{fig:infoHomophily}).
The entropy rates of the ego and alter on a given dyad were correlated (Fig.~\ref{fig:infoHomophily}a).
Figure \ref{fig:infoHomophily}a covers the correlation between the ego and the rank-1 alter (Spearman's $\rho = 0.440$ [0.386, 0.490], $p < 0.001$). 
In Fig.~\ref{fig:infoHomophily}b, we plot the Spearman's $\rho$ between $\hat{h}(\mathrm{ego})$ and $\hat{h}(\mathrm{alter})$ as a function of alter rank. 
These correlations were significant for all ranks ($p < 0.001$).
The correlation drops consistently over the first five or so alters, implying that the homophily effect is connected with contact volume. 
Interestingly, we see weaker associations in cross-entropy (see Supplementary Note 1.8);
further investigation of these and other information homophily effects has the potential to improve our ability to control for homophily in order to explore social contagion.

The ability to repeatedly and accurately predict the text of individuals delivers considerable value to the providers of social media, allowing providers to develop profiles to identify and track individuals\cite{de2013unique,de2015unique} and even manipulate information exposure\cite{pariser2011filter}.
For example, a language model may be trained on available data to generate new text in the ``voice'' of a user\cite{mosteller1963inference,katz1987estimation,bengio2003neural} and with such a language model one could derive a profile for the user by querying it for the likelihood that the user will make certain kinds of statements (for example, how likely are certain statements about one political party or another).
Language models derived in this way can have important consequences: combining predictions from a language model with an algorithm for recommending new social ties, for example, has the potential to create or exacerbate filter bubbles\cite{pariser2011filter}.
The optimal accuracy a trained model can achieve when making these text predictions is mathematically bounded by the predictive information that we estimate here.
That information is so strongly embedded socially underscores the power of the social network: by knowing who the social ties of an individual are and what the activities of those ties are, our results show that one can, in principle, accurately profile even those individuals who are not present in the data\cite{Garciae1701172}. 

Experimental studies are crucial for improving on our results.
For example, we have shown how a platform provider can use information from a user's social ties as a substitute for missing information from that user. 
Yet, in reality, this substituted predictive information can become outdated as the social system and its members are not static entities but evolve over time.
This evolution challenges prediction,
as a user forgoing or deleting their account can change a social tie's future behavior, even if only through the fact of no longer interacting with that user, affecting the future predictability of that user.
Experiments can help understand the effects of this evolution on prediction.
Likewise, any research involving observational data, such as ours, will have difficulty distinguishing social contagion from homophily\cite{aral2009distinguishing,shalizi2011homophily}. 
Our focus on predictive information flow is one of the strongest measures possible given the text data that we study, in that we explicitly utilize the time ordering of information when calculating cross-entropy.
However, information flow alone is not sufficient evidence of contagion.
To establish contagion would require controlling for the tendency of similar alters to share ties,
which the present study provides a first step towards.
Of course, the gold standard for causal influence remains experimental interventions.
Such experimental studies are ideal for studying both dynamic social effects and contagion phenomena.

The time-ordered cross-entropy (Fig.~\ref{fig:introFig}a) applied here to online social activity is a natural, principled information-theoretic measure that incorporates all of the available textual and temporal information.
Although weaker than full causal entailment, by incorporating time ordering, we identify social information flow as the presence of useful, predictive information in the past of one's social tie beyond that of the information in one's own past. 
Doing so closely connects this measure with Granger causality and other strong approaches to information flow\cite{granger1969investigating,schreiber2000measuring}.

\section*{Methods}
\label{sec:methods}

\small
\vspace{-1em}
\singlespacing

\paragraph{Data collection and filtering}

We selected a random sample of individuals for study from the Twitter 10\% Gardenhose feed collected during the first week of April 2014.
From this, we uniformly sampled individuals who had tweeted in English (as reported by Twitter in the metadata for each tweet) during this time period and had 50--500 followers, as reported in the feed metadata. 
The lower follower cutoff is to avoid inactive and bot accounts, whereas the higher cutoff is to ensure that individuals in our sample have comparably sized ego-networks and to avoid studying unusually popular outlier accounts, such as celebrity accounts.
We remark that generating a sample from the 10\% feed necessarily introduces a small bias towards more active individuals,
those who have tweeted at least once into that feed.
For each user, we then collected their complete public tweet history excluding retweets
(up to 3200 most recent public messages, as allowed by Twitter's Public REST API limit\cite{TwitterAPI}).
As discussed later in this section, we then applied to these users a filtering procedure, including both computational tools and human raters to help ensure sufficient data on individual activities and to limit bots and non-individual accounts from our sample.
When finished, we retained a final sample of $n=927$ individual egos and their top-15 alters ($n = 13,905$ total users).
For each initially sampled ego, we collected the user IDs of the account whom the ego `at-mentioned' most frequently in their public tweets, forming the {rank-1 alter}.
Using such mentions is of course not the only way to define a social tie on Twitter;
follower relationships,
numbers of retweets,
or shared textual features (such as hashtags or keywords) could all be reasonably employed to define a social network.
However, defining social ties using mentions gives a stronger signal than simply Twitter following, as it demonstrates active communication on the behalf of at least one of the individuals of a social tie.
Defining social ties as related to the number of mentions also captures a degree of social closeness, whereas follower or following has no strength associated with it.
As was done with the egos, the REST API was then used to retrieve the complete public tweet history of this alter.
Examining the messages of the (ego, rank-1 alter) dyad, we retained egos where the ego's tweets covered at least a 1-year period, the alter's tweets covered at least a 1-year period, and the ego at-mentioned at least 15 unique Twitter users (including the rank-1 alter).
For dyads who satisfied these criteria, we collected the full public messages of the remaining 14 most at-mentioned alters, giving us the full public activities of the ego and their top-15 most mentioned alters.

To limit the effects of bots and non-personal accounts, we moved beyond the basic filtering criteria listed above and employed both computational tools and human raters to examine the accounts of the egos in our dataset. 
These tools were applied in April 2017. 
A small number of accounts in our sample were suspended or deleted after our data collection period and were not available online to be examined, so we simply retained these unrated accounts in our sample.
We used the botometerAPI\cite{BotometerApi,botornot1,botornot2,botornot3,botornot4} to score the probability that an ego account was a bot, and eliminated $n=46$ accounts that scored above 50\%. 
This tool examines Twitter accounts along several dimensions to estimate the likelihood that the account belongs to a bot.
Next, we asked human raters to examine the accounts and report whether the account appeared
to belong to a real person or a non-personal entity, such as a corporation or a bot. Two independent raters examined each account's Twitter homepage if available.
We removed $n=84$ accounts for which both raters agreed that the account did not belong to an individual, beyond those already flagged by the botometer scores.
Raters were recruited on Amazon Mechanical Turk and compensated at a rate of US\$0.10 per three Twitter accounts.
Finally, we also removed a small number of accounts ($n=31$) showing convergence issues with our entropy estimators, as inferred by negative Kullback–Leibler divergences from the ego to the alter or vice versa. 
This gave our final sample size of $n = 927$ egos and their top-15 associated alters, comprising $m=30,852,700$ total tweets.

\paragraph{Control procedures}
We performed two controls for the cross-entropy experiments: random tweets or `temporal control' and random alters or `social control'.
For the temporal control, we constructed proxy tweet streams for the alters that preserved the approximate times at which alters had written messages.
To do this, we substituted for each real alter tweet a randomly sampled English-language tweet posted during the same hour as the real alter tweet.
The randomly sampled replacement tweets were taken from the 10\% Gardenhose feed.
In the social control, we randomized the ego networks, swapping the tweet text streams of true alters with those of randomly chosen alters.
This control does not preserve the times at which the original alters had authored tweets, hence the use of the previous temporal control.

\paragraph{Text processing}

To apply the entropy estimators discussed below, 
we first need to process and tokenize the texts of users.
The UTF-8 encoded text of each user was processed by removing casing, punctuation (except for twitter specific ``@'' and ``\#'' symbols), and URLs (identified as words beginning with ``http://'' or ``https://'').
All tweet texts were concatenated into a single text string in time order (based on the tweet timestamps), except for ``retweets'' which were all excluded in order to focus on the effect of shared language and avoid artificially inflating predictability scores. 
The text was then tokenized into words by segmenting on whitespace.

\paragraph{Measuring information in written text}

The entropy (rate) $h$ of a sequence of words is the number of bits needed to
encode the next word, on average, given past words.
Kontoyianni \emph{et al.}\cite{kontoyiannis_nonparametric_1998} proved convergence for a nonparametric estimator $\hat{h}$ for $h$:
\begin{equation}
    \hat{h} = \frac{N \log N}{\sum_{i=1}^{N} \Lambda_{i}},
    \label{eqn:hhat}
\end{equation}
where $N$ is the length of the sequence of words and $\Lambda_{i}$ is the match length of the prefix at position $i$, that is, it is the length of the shortest subsequence (of words) starting at $i$ that has not previously appeared. 
(All logarithms are base 2.)
If the sequence of words were randomly shuffled, breaking any long-range structure, this estimator converges to the traditional Shannon entropy on unigrams (see Supplementary Fig.~1).

The ideas underlying estimators such as Eq.~\eqref{eqn:hhat} play an important role in the mathematics of data compression algorithms. 
Indeed, some authors have used practical compression software to estimate the information content of a text.
However, such estimates tend to be biased, as specific compression implementations (such as gzip) tend to sacrifice small amounts of extra compression to run much more efficiently. 
Owing to these approximations, it is important to work directly with the theoretical estimator to more accurately estimate $h$, as we have when we applied Eq.~\eqref{eqn:hhat}.

\paragraph{Measuring the flow of predictive information}

To generalize Eq.~\eqref{eqn:hhat} to a cross-entropy between two sequences $A$ and $B$, we define the \textbf{cross-parsed match length} $\Lambda_{i}(A | B)$ as the length of the shortest subsequence starting at position $i$ of sequence $A$ not previously seen in sequence $B$. 
If sequences $A$ and $B$ are \textbf{\textit{time-aligned}}, as in timestamped social media posts, then `previously' refers to all of the words of $B$ written prior to $t_{i}(A)$, the time when the $i$th word of $A$ was posted, according to the timestamp of the respective tweet. 
The estimator for the cross-entropy rate is then
\begin{equation}
	\hat{h}_{\times}(A \mid B) = \frac{N_{A} \log N_{B}}{\sum_{i=1}^{N_{A}} \Lambda_{i}(A \mid B)},
	\label{eqn:crossEntropy}
\end{equation}
where $N_{A}$ and $N_{B}$ are the lengths of $A$ and $B$, respectively. 
An estimator of the relative entropy (or KL-divergence),  similar to Eq.~\eqref{eqn:crossEntropy}, was introduced by Ziv and Merhav\cite{243444}.
The log term in Eq.~\eqref{eqn:crossEntropy} has changed to $\log N_{B}$ because now $B$ is the `database' (or window, in Lempel-Ziv terms) we are searching over when we compute the match lengths; the $N_{A}$ factor is due to the average of the $\Lambda_{i}$'s taking place over $A$. 
The cross-entropy tells us how many bits on average we need to encode the next word of $A$ given the information previously seen in $B$. 
Furthermore, $\hat{h}_{\times}(A \mid A) = \hat{h}$.
The cross-entropy can be applied directly to an ego-alter pair by choosing $B$ to be the text stream of the alter and $A$ the text stream of the ego.

We now wish to generalize the cross-entropy to $\hat{h}_{\times}(A \mid \mathcal{B})$, estimating the average amount of information needed to encode the next word of sequence $A$ given the information in a \emph{set} of sequences $\mathcal{B}$.
Take the cross-parsed match length for a set of databases to be $\Lambda_{i}(A \mid \mathcal{B}) = \max\{ \Lambda_{i}(A \mid B), B \in \mathcal{B} \}$, that is, the longest match length over any of the sequences in $\mathcal{B}$.
This cross-parsing implies a new $\log N_{A\mathcal{B}}$ factor in the estimator, where $N_{A\mathcal{B}}$ is the average of the lengths $N_{B}$ ($B \in \mathcal{B}$), weighted by the number of times matches were found in each sequence $B \in \mathcal{B}$. (If the same match length occurs for more than one sequence $B \in \mathcal{B}$ then each such sequence receives a weight in the average.) The estimator is
\begin{equation}
	\hat{h}_{\times}(A \mid \mathcal{B}) = \frac{N_{A} \log N_{A\mathcal{B}}}{\sum_{i=1}^{N_{A}} \Lambda_{i}(A \mid \mathcal{B})},
	\label{eqn:jointCrossEntropy}
\end{equation}
where
	$N_{A\mathcal{B}} = \left.{\sum_{B \in \mathcal{B}} w_{B} N_{B}}\middle/{\sum_{B \in \mathcal{B}} w_{B}}\right.$
and $w_{B}$ is the number of times that matches from $A$ are found in $B \in \mathcal{B}$.
Note that $\sum_{B} w_{B} \geq N_{A}$ due to possible ties, with equality holding if no ties occur.
Note that Eq.~\eqref{eqn:jointCrossEntropy} reduces to Eq.~\eqref{eqn:crossEntropy} when $\left|\mathcal{B}\right| = 1$.

Equation \eqref{eqn:jointCrossEntropy} lets us build the cumulative cross-entropy by appropriate choices of $\mathcal{B}$. 
Here, we sequentially added alters to the set $\mathcal{B}$ in order of decreasing contact volume (i.e., $\mathcal{B} = \{\mathrm{alters}\}$), to understand how information grows as more alters are made available. Likewise, Eq.~\eqref{eqn:jointCrossEntropy} lets us build the transfer entropy-like measures by additionally including the ego within the set $\mathcal{B}$ (i.e., $\mathcal{B} = \{\mathrm{ego}\} \cup \{\mathrm{alters}\}$).

We implemented Eqs.~\eqref{eqn:hhat}-\eqref{eqn:jointCrossEntropy} in Python.
See code availability statement.

\paragraph{Estimator convergence on our data}

The estimator given by Eq.~\eqref{eqn:hhat} has been proven to converge asymptotically under stationarity assumptions\cite{kontoyiannis_nonparametric_1998}. However, our data are finite, and so we investigated the convergence properties of the estimator empirically (see Supplementary Figure 1b,c).
In general, we observed that the entropy \eqref{eqn:hhat} saturates after around 1000 tweets (approximately 10,000 words).
Likewise, the cross-entropy estimator $h_\times(A\mid B)$ tends to converge within around 50\% of the ego's observed lifespan (see Supplementary Note 1.1).

\subsection*{Code Availability}
The code used to generate the results of this paper is available from the corresponding authors upon request.

\subsection*{Data Availability}
Data that support the findings of this study are available at \href{http://dx.doi.org/10.6084/m9.figshare.7338992}{Figshare}.

\vspace{0.75\baselineskip}

\renewcommand\refname{References}

\subsection*{Acknowledgements}
We gratefully acknowledge the resources provided by the Vermont Advanced Computing Core.
This material is based upon work supported by the National Science Foundation under Grant No.\ IIS-1447634 (J.P.B.).
L.M.\ acknowledges support from the Data To Decisions Cooperative Research Centre (D2D CRC), and the ARC Centre of Excellence for Mathematical and Statistical Frontiers (ACEMS).
The funders had no role in study design, data collection and analysis, decision to publish or preparation of the manuscript.

 \subsection*{Author Contributions}
J.P.B.\ and L.M.\ designed the research. L.M.\ oversaw data collection and processing.
X.L.\ collected and analyzed human rater data. J.P.B.\ and L.M.\ analysed the data and wrote the manuscript.

 \subsection*{Competing Interests} The authors declare that they have no competing financial interests.

\paragraph{Correspondence and requests for materials} should be addressed to J.P.B.\ (email:
 \href{mailto:james.bagrow@uvm.edu}{james.bagrow@uvm.edu}) or L.M.\ (email:
 \href{mailto:lewis.mitchell@adelaide.edu.au}{lewis.mitchell@adelaide.edu.au}).

\clearpage{}


\section*{Supplementary material}

# Contents

# List of Supplementary Figures

# List of Supplementary Tables

# Supplementary Note 1

## Cross-entropy estimator convergence

Supplementary Figure 1 shows the difference between our entropy estimator and traditional Shannon entropy (panel a), as well as the estimator convergence (panels b & c). See the Methods section in the main text for more information.

1

For the cross-entropy estimator $h_\times(A \mid B)$, we examined the convergence over the lifespan or time window within which the ego has authored tweets. Supplementary Figure 2 shows the convergence of the cross-entropy for the rank-1 alter $h_\times(\text{ego}|\text{alter 1})$ (left panel), where we truncate both the ego and alter's tweets after some fraction of the ego's lifespan. In general, we found that the cross-entropy estimator saturates within around 50% of the ego's lifespan. The right panel shows a histogram of the slopes of the convergence curves for all users over the final 25% of the ego's lifespan, as a fraction of the final value of $h_\times(\text{ego} \mid \text{alter 1})$. These slopes were computed via linear regressions, and many of the slopes are close to zero.

The cross-entropy can also be associated with the predictability by applying Fano's Inequality [1]. Fano's Inequality relies on both the entropy and the cardinality of the random variable; here we take the size of the ego's unique vocabulary as this is the variable we are trying to predict. In Supplementary Figure 3 we present the relationship between cross-entropy and predictability for our data compared with solid lines denoting constant vocabulary-size curves. The predictability of the ego given the alter is lower than the predictability of the ego given the ego because the cross-entropy is greater than the entropy, capturing the increased uncertainty (decreased information) we have by trying to predict the ego given the alter instead of the ego.

## Supplementary Note 2

### Extrapolating cross-entropy and predictability

We are limited by our data to a window of the top-15 most frequently contacted alters per ego. To address a limit of entropy or predictability as more alters are added, we used a saturating function to extrapolate beyond alter rank $r = 15$.

Specifically, we extrapolated the cross-entropy using the function

$$h(r) = h_\infty + \frac{\beta_0}{\beta_1 + r}, \tag{S1}$$

with the goal of identifying the value of $h_\infty$ and, perhaps more realistically, to estimate $h(r_{\text{dunbar}})$, where $r_{\text{dunbar}} \approx 150$ [2]. Using Levenberg-Marquardt for nonlinear regression, we found best fit parameters of (value ± 95% CI):

$$h_\infty = 5.761978 \pm 0.089699,$$
$$\beta_0 = 9.455984 \pm 1.358027,$$
$$\beta_1 = 2.553345 \pm 0.444479,$$

for the cross-entropy of the ego given the alters.

In Supplementary Figure 4 we show the mean cross-entropy as a function of alter rank and compare it with the results of the fitted function. The fit is reasonable. Similarly, fits of the same functional form were applied to the predictability (ego given alter) curves:

$$\Pi(r) = \Pi_\infty + \frac{\beta_0}{\beta_1 + r}, \tag{S2}$$

and here we found best fit parameters

$$\Pi_\infty = 0.608219 \pm 0.006914,$$
$$\beta_0 = -0.734410 \pm 0.100195,$$
$$\beta_1 = 2.320039 \pm 0.398486.$$

We also experimented with a second form of extrapolating function:

$$h(r) = h_\infty + \beta_0 r^{-\beta_1}, \qquad \Pi(r) = \Pi_\infty + \beta_0 r^{-\beta_1}. \tag{S3}$$

This function, referred to as **Function 2**, also fits the data well (Supplementary Figure 5) but is a bit less conservative in its extrapolation prediction when extrapolating for $r \to \infty$. To further compare Function 2 and the original function (Function 1), we plotted the residuals between the fits and the data in Supplementary Figure 5.

Taken together, we see that Function 1 (Eqs. (S1) and (S2)), the more conservative estimator, has consistently smaller residuals than Function 2. Both functions' residuals were statistically independent of the exogenous variable $r$ ($p > 0.05$). We concluded that Function 1 is a better choice since it has smaller residuals and is more conservative than Function 2.

## Supplementary Note 3

### Vocabulary sizes on social media

In Supplementary Figure 6 we present the distributions of the total number of words written per ego and the number of unique words (the vocabulary size) per ego, for the users in our dataset. Egos had (mean ± s.d.) $26802.76 \pm 9061.53$[1] total numbers of words and $5207.44 \pm 1769.48$ numbers of unique words. The latter quantity, the vocabulary size, was used in Fano's Inequality to compute the predictability.

## Supplementary Note 4

### Information content on social media compared with formal written text

To contextualize the entropy rates we estimated for our dataset (most egos had entropies of $5.5 < h < 8$ bits), we compared the entropy rates of formal text with the rates of Twitter users to better understand the information content of social media writings[2]. First, we considered the entropies of some famous example texts (Supplementary Table 1). We considered writers who were famous for being very simple in style (Hemingway) and very complex (Joyce) and found this was reflected in the entropy rates (5.87 bits for Hemingway compared with 7.06 bits for Joyce). The higher entropies reflect that Joyce's word choices are less regular and less predictable than Hemingway's. These formally written and edited texts are very different from social media posts, and yet the range of entropies values we observed was compatible to some extent.

We also took the standard ***Brown corpus*** [4], a benchmark text set used in natural language processing and computational linguistics research, as a large-scale baseline of formal text. The corpus consists of approximately 1M words and covers 500 writing samples across 15 fiction and non-fiction categories. Each

---

[1]For context, this is about the typical length of a novella, defined by the Science Fiction and Fantasy Writers of America as 17500–39999 words [3].

[2]The texts were processed by removing punctuation and casing.

category was broken into 10-thousand-word chunks and the entropies of these chunks were computed. Individual chunks did not span multiple categories and if a chunk at the end of a category was less than 10 thousand words it was discarded to ensure all entropy estimates were computed using the same amount of data. This gave $n = 93$ samples.

We found that formal and social text have the same average value but that the variation across the Twitter sample was significantly greater than across the formal text (Supplementary Figure 7).

## Supplementary Note 5

### A censoring filter to determine long-range information in the egos and alters

To study the recency of information we applied the (cross-)entropy estimators to censored text, where we removed the recent past of the text and asked how much if any information is lost. If most predictive information is in the recent past, by removing it we should see a significant change in the cross-entropy, although there should always be some loss in information, as the sequences being matched across are always getting shorter.

Specifically, to compute the original cross-entropy (Eq. (2)) between two sequences $A$ and $B$, we need the cross-parsed match length at position $i$, $\Lambda_i(A \mid B)$, giving us the shortest subsequence of words in $A$ beginning at position $i$ not seen previously in $B$. This last part, the past of $B$, is based on the timestamps of the words: we search all words in $B$ written before the $i$th word $w_i$ in $A$: $[w_j \in B \mid t_j(B) < t_i(A)]$, where $t_i(A)$ is the time when the $i$th word in $A$ was posted (taking all words in a single tweet to be posted at the time the tweet was posted).

The censoring filter simply truncates the past of $B$ at each position $i$. Instead of searching all of the past of $B$ we instead search the past older than an amount $\Delta T$: $[w_j \in B \mid t_j(B) < t_i(A) - \Delta T]$. By censoring $B$ as we sweep forward in the computation of the cross-entropy, we can estimate how much information is recent versus long-term on average by the change in the cross-entropy rate as a function of $\Delta T$. The same calculation holds for the "self" entropy, simply by setting $B = A$.

We measured the loss of information in the main text out to 24 hours. Here we complement that calculation with Supplementary Figure 8 which presents the information loss out to 1 week. We see in both curves that long-range information is lost by the increasing trend. However, the trend is more shallow for the alters than the ego: taking away more of the ego's past removes more information about the ego than taking away the less recent pasts of the alters.

This censoring filter reduces the amount of data available from which to compute the cross-parsed match lengths $\Lambda_i(A \mid B)$. We investigate the extent of this data loss in Supplementary Figure 9, which shows how the number of censored alter tweets per ego tweet $n_{\text{cens}}$ depends on the censoring interval $\Delta T$. The left panel shows distributions of the mean $n_{\text{cens}}$ for different lags $\Delta T$ from 0.5 to 6 hours. Each distribution is obtained by counting the number of tweets posted by all alters in $\Delta T$ hours before each ego tweet, taking the mean of these values, and then plotting the distribution of these mean values. The right panel shows the mean and 95% quantiles of these distributions as a function of $\Delta T$. As expected, the mean $n_{\text{cens}}$ increases roughly linearly as a function of $\Delta T$, however the numbers of censored tweets remain relatively low compared to the total amount of data available ($\approx 3200$ tweets for each ego).

4

# Supplementary Note 6

## Posting frequency and predictability

Main text Fig. 2b demonstrated associations between predictability of the ego given the alter and how frequently either the ego or the alter posted to Twitter, as quantified by the average number of posts per day. Egos who posted more than 8 times per day on average were 16.5% ± 14.9% (difference in mean predictability ± 95% CI on mean) more predictability from their alter than egos who posted less than 1 time per day on average. Likewise, egos where the alter posted more than 8 times per day on average were 23.8% ± 4.46% less predictability from the alter than egos where the alter posted less than 1 time per day on average. These changes in predictability show that egos who post more frequently are more predictable from their alters than less frequent posting egos, while the opposite association holds about egos with frequent and infrequent posting alters. However, these differences in predictability only highlight the extreme ends of the data range, so we also measured the statistical association across the entire posts-per-day range: all measured associations were significant (Supplementary Table 2).

Further, in Supplementary Figures 10 and 11 the association between the posting frequencies of the egos and alters with the predictability of the ego (Supplementary Figure 10: top row), predictability of the alter (Supplementary Figure 10: bottom row), predictability of the ego given the alter (Supplementary Figure 11: top row), and the predictability of the alter given the ego (Supplementary Figure 10: bottom row). We found that the predictabilities of the egos and alters are roughly independent of their posting frequency, except for very infrequently posting alters (Supplementary Figure 11: bottom right), which is likely a result of insufficient data.

The associations between posting frequency and the cross-predictability of the ego given the alter hold even when considering all alters not just the rank-1 alter, as we did in the main text (Supplementary Figure 11: top row). Likewise, the trends also hold (in reverse) when considering the predictability of the alter given the ego (Supplementary Figure 11: bottom row).

# Supplementary Note 7

## Contact volumes and predictability

Here we present in Supplementary Figure 12 the predictability across social ties as a function of how often those social ties contact one another. We ranked the ties of individuals in descending order. Working with ranks helps to account for the variability in contact volumes and overall activity levels across users of social media. Across ranks we found a significant decrease in predictive information, in both directions (predicting the ego given the alter and predicting the alter given the ego).

# Supplementary Note 8

## Cross-entropy homophily

In the main text we reported a homophily between the entropies of the egos and their alters. Here we explore a similar association with their cross-entropies.

The cross-entropies between the egos and alters are less well correlated, either with the cross-entropy in

5

the opposite direction, or with the entropies themselves. The correlations (for the rank-1 alters) are:

$$R\left(\hat{h}(\text{ego}), \hat{h}(\text{alter})\right) = 0.478,$$

$$R\left(\hat{h}_\times(\text{ego} \mid \text{alter}), \hat{h}_\times(\text{alter} \mid \text{ego})\right) = -0.122,$$

$$R\left(\hat{h}(\text{ego}), \hat{h}_\times(\text{ego} \mid \text{alter})\right) = 0.240,$$

$$R\left(\hat{h}(\text{ego}), \hat{h}_\times(\text{alter} \mid \text{ego})\right) = 0.227,$$

$$R\left(\hat{h}(\text{alter}), \hat{h}_\times(\text{ego} \mid \text{alter})\right) = 0.247,$$

$$R\left(\hat{h}(\text{alter}), \hat{h}_\times(\text{alter} \mid \text{ego})\right) = 0.300.$$

While significant in all cases, the correlations between the (self) entropies $\hat{h}(\text{ego})$ and $\hat{h}(\text{alter})$ are stronger than between any of the cross-entropies, demonstrating that the effects captured by the cross-entropies over a dyad are different than that captured by the entropies of the individuals in that dyad.

## Supplementary Note 9

### Reciprocity and information flow

Table 3 provides statistical analyses of the associations reported in main text Fig. 3. Due to potential nonlinear relationships, we report monotonicity coefficients (both Spearman's Rho and Kendall's Tau). We see significant associations between predictability of the ego given the alter and the number of social ties of the ego (cf. main text Fig. 3a). Likewise, we see significant associations between contact volume and predictability of the ego given the alter, with a positive trend for alter-mentions-ego contact volume and a slight negative trend for ego-mentions-alter contact volume. This asymmetry supports information flow capturing directionality in relationships (cf. main text Fig. 3b).

In the main text we reported on the relationship between contact volume and information flow, as measured by the cross-entropy (and mapped into the predictability. A closely related quantity often employed in this context is the Kullback-Leibler divergence, or KL-divergence, $KL(\text{ego} \parallel \text{alter}) \equiv \hat{h}_\times(\text{ego} \mid \text{alter}) - \hat{h}(\text{ego})$ [1]. In our data the correlation between $\hat{h}_\times(\text{ego} \mid \text{alter})$ and $KL(\text{ego} \parallel \text{alter})$ is quite high (Supplementary Table 4) and so they are effectively the same measure.

We showed that alters who more frequently mention their ego provide more predictive information (lower cross-entropy/KL-divergence) than alters who less frequently mention their ego. Meanwhile, the converse was not true: the ego can mention the alter more or less, and there was not an association with the predictive information possessed by the alter about the ego.

Here we supplement that result by reversing the perspective—instead of asking about the predictive information about the ego possessed by the alter we ask about the predictive information about the alter possessed by the ego. We measure this with the reversed KL-divergence, $KL(\text{alter} \parallel \text{ego}) \equiv \hat{h}_\times(\text{alter} \mid \text{ego}) - \hat{h}(\text{alter})$. With this reversal we should expect to also see a reversal in the association of contact volume, and we found this to be the case (Supplementary Figure 13). In Supplementary Figure 13 we compared both KL-divergences and saw that the trends approximately reverse, as expected.

# Supplementary Note 10

**Interrelations between information-theoretic quantities**

In Supplementary Table 4 we present the Spearman rank correlation coefficients between the primary information-theoretic quantities we computed, including the KL-divergence: $KL(\text{ego} \parallel \text{ater}) \equiv \hat{h}_{\times}(\text{ego} \mid \text{alter}) - \hat{h}(\text{ego})$. The cross-entropy and KL-divergence are strongly correlated.

# Supplementary Figures

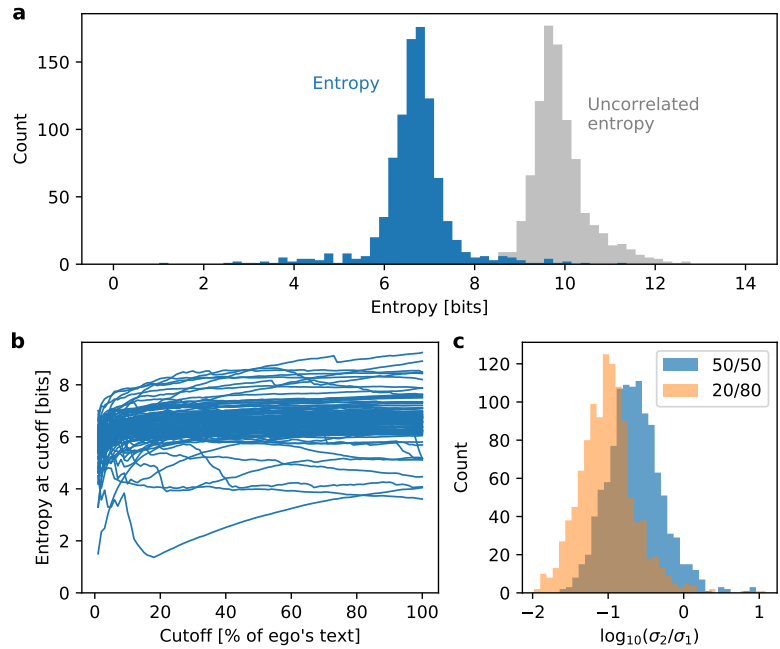

Supplementary Figure 1: **a**, Correlations in the text account for ≈3 additional bits of information. The uncorrelated entropy (considering only the relative frequencies of words posted by Twitter users) is approximately 3 bits higher than the correlated entropy as estimated from Eq. 1. **b & c**, Entropy estimator convergence. **b**, The estimator generally saturates well within our data window, as evidenced by the flattening of the entropy estimate as we examine more of the ego's text. **c**, Here we compute the variance of each ego's entropy over two portions of the curves at left. One distribution compares the variance of the final 50% of the data to the initial 50%, while the other compares the variance of the final 20% of the data to the initial 80%. The latter shows significantly smaller variability, underscoring how entropy estimates have converged within our data window. The left plot shows a random selection of egos, while the right covers all dyads in our dataset.

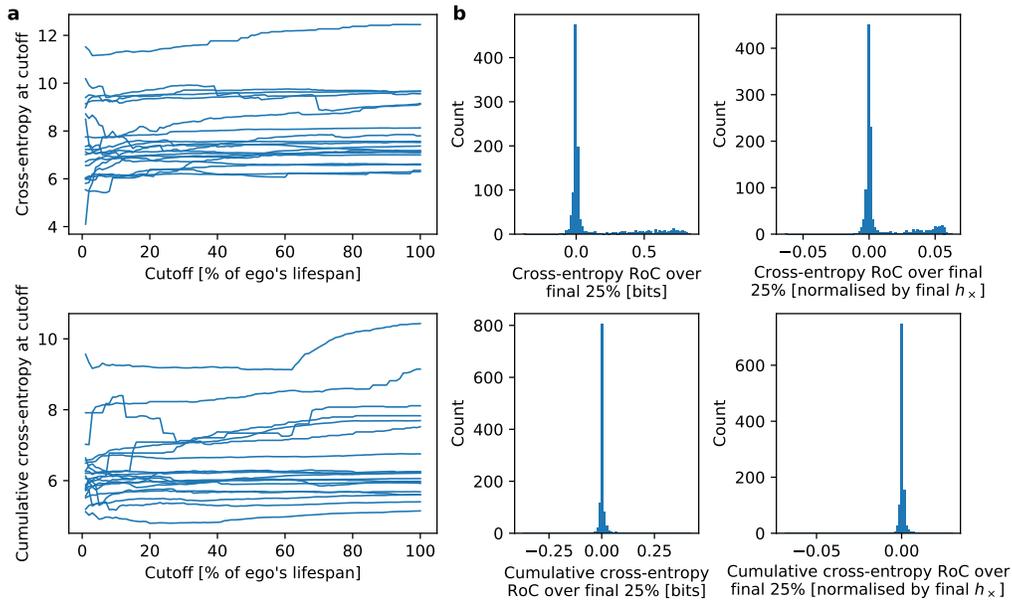

Supplementary Figure 2: Convergence of the cross-entropy estimator. **a**, The estimator saturates well within the lifespan of the ego's tweets, generally within 50% of the lifespan. **b**, The distributions of the slope (RoC: rate-of-change) over the final 25% of the curves. The majority of egos have very flat RoCs at the end of their data windows. In the left plots we show a randomly selection of egos, while the distributions on the right curve all dyads in our dataset.

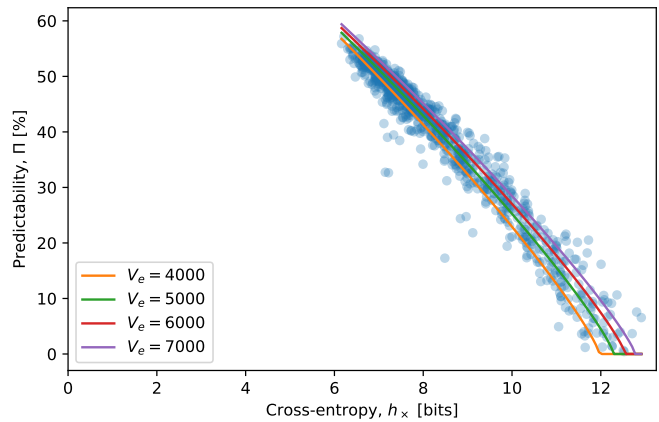

Supplementary Figure 3: Cross-entropy $\hat{h}_\times(\text{ego} \mid \text{alter})$ and predictability $\Pi$ across different ego vocabulary sizes $v_e$ indicated by color.

9

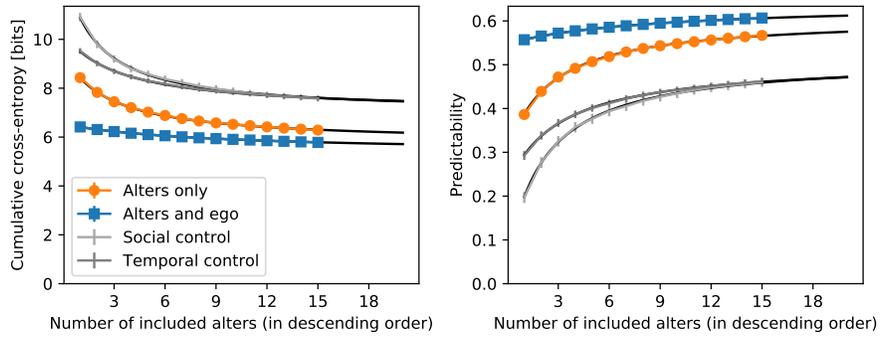

Supplementary Figure 4: Extrapolating cross-entropy and predictability. The fitted functions (Eqs. (S1) and (S2), solid lines) compared with the original cross-entropy data (averaged for each alter rank). Note that the function was fitted to the original and not averaged cross-entropy values. Points and error bars denote means and 95% CIs, respectively, on the data, while error bars without points denote the same quantities for the social and temporal controls. Colors distinguish the controls (light gray: social control; dark gray: temporal control), and whether the ego's information was included alongside the alters' (blue: alter and ego; orange: alter only). After fitting to the original data, the extrapolating function was plotted out to a value of 20 (beyond our data window of 15 alters) to highlight the extrapolation.

10

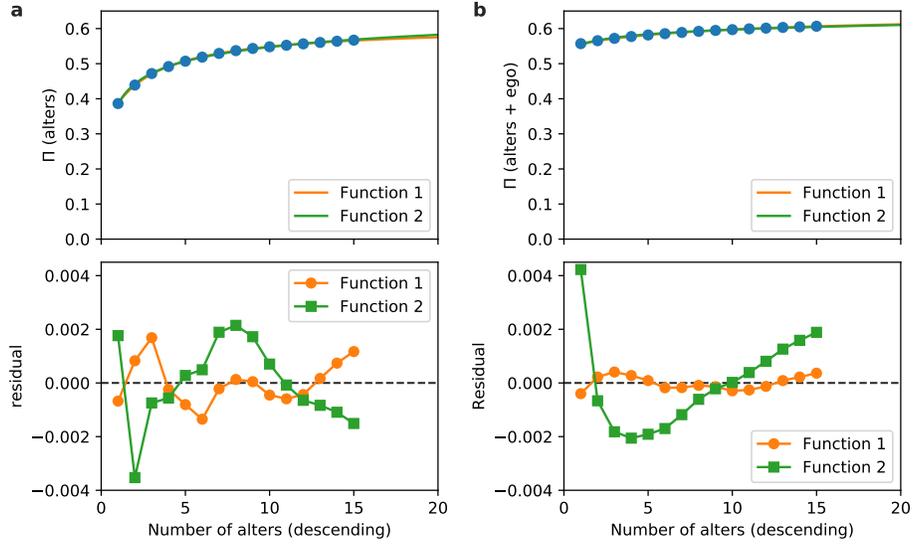

Supplementary Figure 5: Extrapolations and residuals for the predictability functions (Function 1: Eq. (S2); Function 2: Eq. (S3)). **a**, Comparison of the measured cross-entropies (for the top-15 alters) with the extrapolation functions and mean residuals between function fit and original data. **b**, Same as panel A but for fits of the same form as Eqs. (S2) and (S3) to $\Pi(r)$ including the past of the ego along with the alters. Overall, Function 1 was slightly more conservative than Function 2, extrapolating to a slightly smaller value of $\Pi$, and had lower residuals. Colors denote the data (blue), Function 1's fit and residual (orange), and Function 2's fit and residual (green).

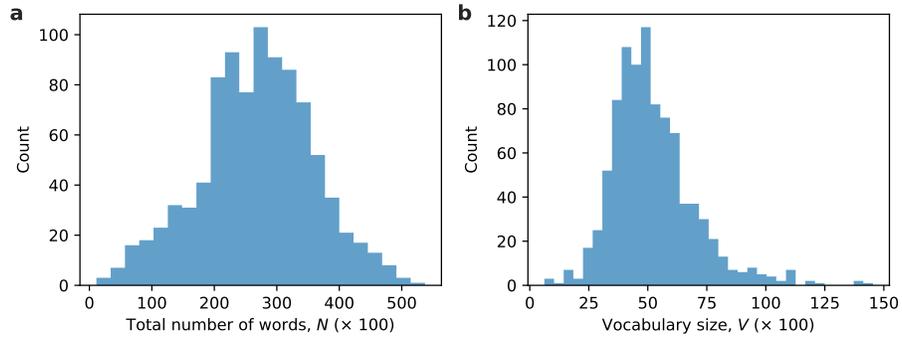

Supplementary Figure 6: Distributions of Twitter ego vocabulary size. **a**, The total number of words written. **b**, The vocabulary size or number of unique words written.

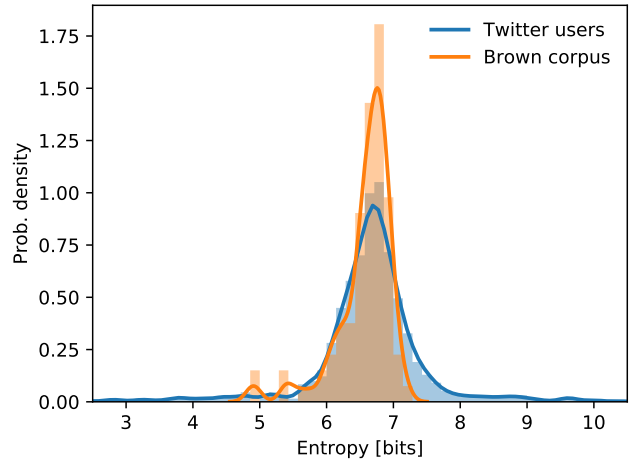

Supplementary Figure 7: Entropy distributions for social and formal written text corpora. We found that the distributions have the same central tendency (Mann-Whitney U test: $U = 39797$, $p > 0.1$) but different dispersions (Fligner-Killeen test on variances, $\chi^2 = 15.580$, $p < 10^{-4}$ ) Brown corpus was taken from NLTK v3.2.1 corpora [5].

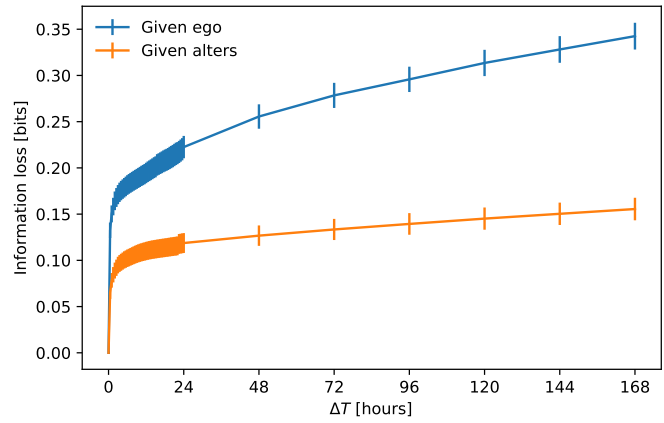

Supplementary Figure 8: Alters provided less long-range information about the ego than the ego itself. This plot complements the loss in predictability shown in the main text and extends $\Delta T$ beyond the 24-hour window to a one-week period. Error bars show 95% CIs. Up to 24 hours we show information loss every 30 minutes; every 24 hours thereafter.

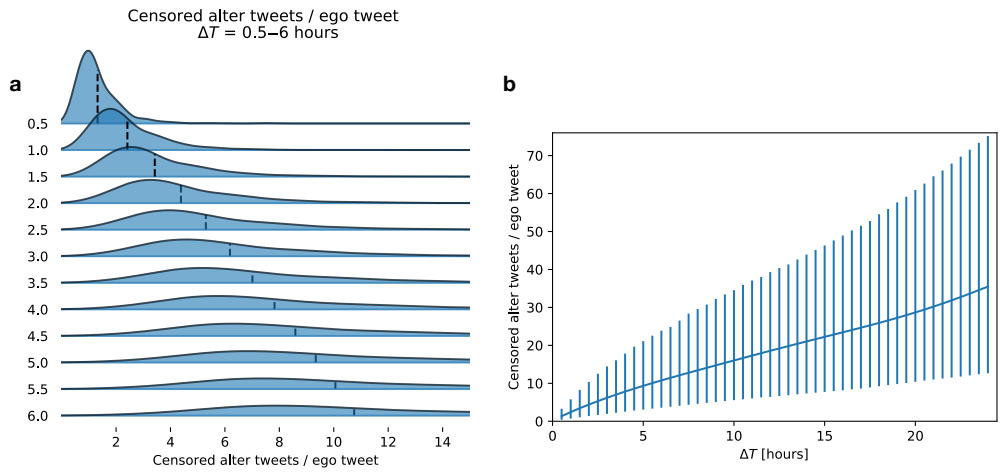

Supplementary Figure 9: Number of alter tweets censored per ego tweet increases with $\Delta T$. **a**, Distributions of mean number of censored alter tweets per ego tweet for lags $\Delta T$ from 0.5–6 hours. Vertical lines show the mean of each distribution. **b**, Mean number of censored alter tweets/ego tweet as a function of $\Delta T$. Error bars show 95% quantiles.

15

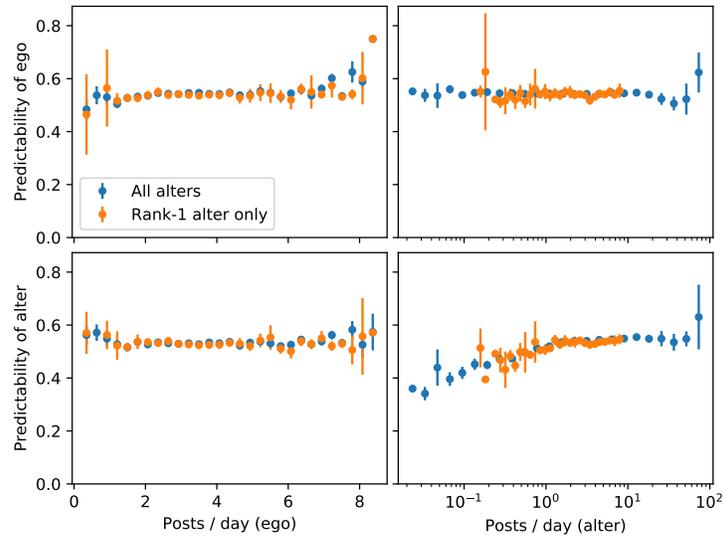

Supplementary Figure 10: Self-predictabilities are approximately independent of activity frequency, with the exception of predictability of the alter as a function of the alter's activity frequency (lower right). This is primarily due to insufficient data: alters who post very infrequently have low predictability, but qualitatively the trend levels off for alters who post more than ≈1 time per day. Error bars show 95% CIs.

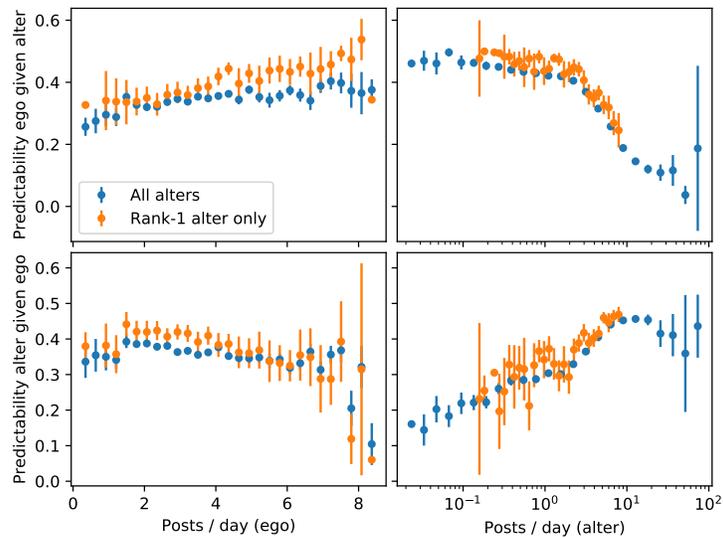

Supplementary Figure 11: Association between cross-predictability and posting frequency holds for all alters. Here we repeated the trends shown in the main text where we considered the rank-1 alter only, but now we included all alters as well. Due to alters who post very frequently and very infrequently, we used a logarithmic scale on the right column. Error bars show 95% CIs.

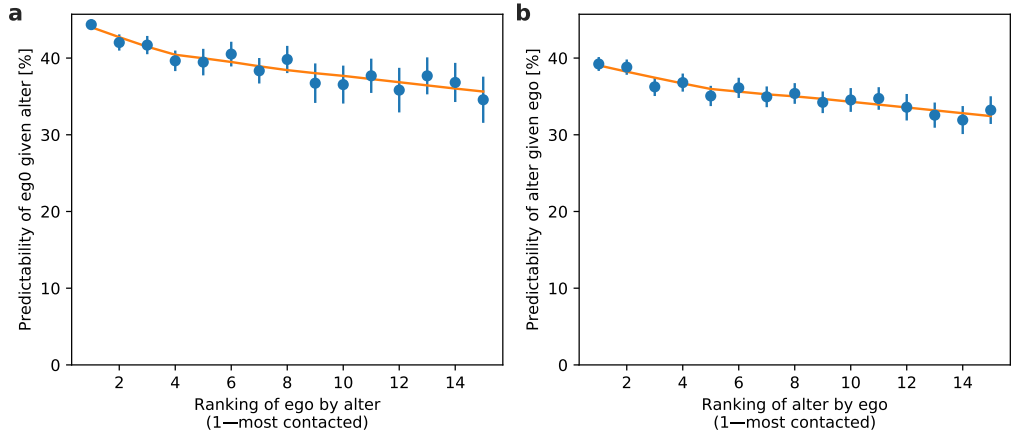

Supplementary Figure 12: Less frequently contacted ties provide less predictive information. Here we plot the mean predictability of the ego given the alter averaged over ego-alter pairs conditioned on the rank of the ego by the alter (panel **a**) or rank of the alter by the ego (panel **b**), with rank-1 being the most frequently contacted social tie. Error bars show 95% CIs and the solid line denotes a LOWESS fit that provides a guide for the eye.

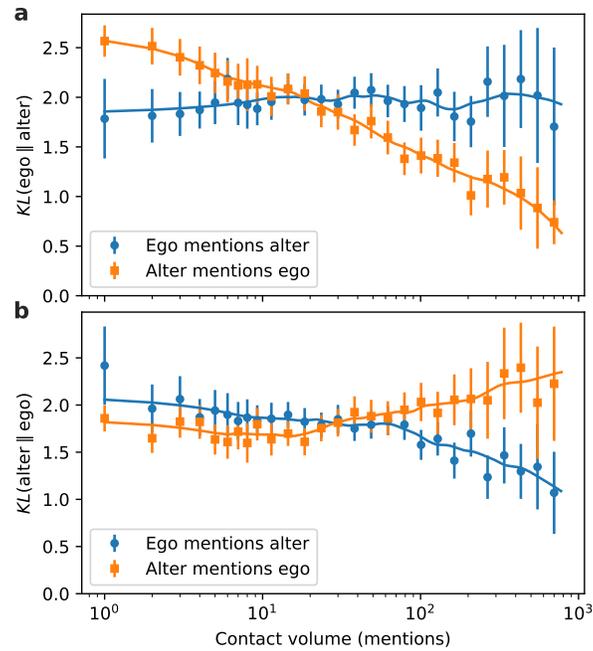

Supplementary Figure 13: Reciprocated information flows are captured in both directions. **a**, In the main text we reported on the trend between contact volume and the cross-entropy from the alter to the ego. We repeat that figure here but with the KL-divergence. **b**, In comparison, if we consider the opposite divergence, from the ego to the alter, we see a similar trend but reversed: egos which more frequently mention their alter give more predictive information (lower divergence) than egos which mention their alter less often. In both panels, error bars show 95% CIs, solid lines denote LOWESS fits that provide a guide for the eye, and colors indicate direction of contact (either ego mentions alter or alter mentions ego).

# Supplementary Tables

Supplementary Table 1: Entropy rates of some example texts. Samples 1 and 2 were the first and second 38,000 words of each text, respectively (a bit longer than the typical Twitter user's text stream). Hemingway is known for his simple writing style while Joyce is famous for the opposite; this is well reflected in their respective entropy rates.

| Text | Author | $\hat{h}$ (sample 1) [bits] | $\hat{h}$ (sample 2) [bits] |
|---|---|---|---|
| For Whom the Bell Tolls | Ernest Hemingway | 5.870953 | 5.910003 |
| Gravity's Rainbow | Thomas Pynchon | 5.881336 | 5.881336 |
| The Fellowship of the Ring | J.R.R. Tolkien | 6.439215 | 6.340354 |
| Ulysses | James Joyce | 7.067339 | 7.227677 |

Supplementary Table 2: Predictability vs. posting frequency (cf. main text Fig. 2b). Both trends shown in Fig. 2b had statistically significant correlations, reported here using the Spearman's rho and Kendall's tau correlation measures. Confidence intervals on $\tau$ were computed as per Sec. 8.3 of Hollander *et al.* [6].

|  | Spearman's $\rho$ [95% CI] | p-value | Kendall's $\tau$ [95% CI] | p-value |
|---|---|---|---|---|
| Posts / day (ego) | 0.276 [0.216, 0.335] | $< 10^{-16}$ | 0.183 [0.145, 0.222] | $< 10^{-16}$ |
| Posts / day (alter) | −0.437 [-0.487, -0.383] | $< 10^{-43}$ | −0.291 [-0.327, -0.256] | $< 10^{-39}$ |

Supplementary Table 3: Predictability vs. social ties (cf. main text Fig. 3a) and contact volume (cf. main text Fig. 3b). The asymmetry in the associations of predictability with the two directions of contact volume demonstrate how information flow captures the directionality of relationships. Confidence intervals on $\tau$ were computed as per Sec. 8.3 of Hollander *et al.* [6].

|  | Spearman's $\rho$ [95% CI] | p-value | Kendall's $\tau$ [95% CI] | p-value |
|---|---|---|---|---|
| Num. social ties of alter | −0.199 [-0.224, -0.175] | $< 10^{-53}$ | −0.133 [-0.150, -0.117] | $< 10^{-52}$ |
| Contact vol. (Ego → alter) | −0.0185 [-0.0440, 0.00704] | 0.156 | −0.0124 [-0.0290, 0.00500] | 0.156 |
| Contact vol. (Alter → ego) | 0.226 [0.202, 0.250] | $< 10^{-68}$ | 0.154 [0.137, 0.170] | $< 10^{-67}$ |

Supplementary Table 4: Cross-entropy and KL-divergence are strongly correlated (bold). Here we present the nonparametric Spearman correlation between information-theoretic quantities computed over the $n = 927$ ego-(rank-1 alter) dyads.

|  | $\hat{h}$(ego) | $\hat{h}$(alter) | $\hat{h}_\times$(e ∣ a) | $\hat{h}_\times$(a ∣ e) | $KL$(e ∥ a) | $KL$(a ∥ e) |
|---|---|---|---|---|---|---|
| $\hat{h}$(ego) | 1.000000 | 0.439867 | 0.302718 | 0.201535 | -0.142122 | 0.000338 |
| $\hat{h}$(alter) | 0.439867 | 1.000000 | 0.257699 | 0.236107 | 0.048474 | -0.176895 |
| $\hat{h}_\times$(e ∣ a) | 0.302718 | 0.257699 | 1.000000 | -0.281961 | **0.858274** | -0.395016 |
| $\hat{h}_\times$(a ∣ e) | 0.201535 | 0.236107 | -0.281961 | 1.000000 | -0.384678 | **0.881891** |
| $KL$(e ∥ a) | -0.142122 | 0.048474 | **0.858274** | -0.384678 | 1.000000 | -0.409757 |
| $KL$(a ∥ e) | 0.000338 | -0.176895 | -0.395016 | **0.881891** | -0.409757 | 1.000000 |




\begin{thebibliography}{10}
\expandafter\ifx\csname url\endcsname\relax
  \def\url#1{\texttt{#1}}\fi
\expandafter\ifx\csname urlprefix\endcsname\relax\def\urlprefix{URL }\fi
\providecommand{\bibinfo}[2]{#2}
\providecommand{\eprint}[2][]{\url{#2}}

\bibitem{Kossinets2006}
\bibinfo{author}{Kossinets, G.} \& \bibinfo{author}{Watts, D.~J.}
\newblock \bibinfo{title}{{Empirical Analysis of an Evolving Social Network}}.
\newblock \emph{\bibinfo{journal}{Science}} \textbf{\bibinfo{volume}{311}},
  \bibinfo{pages}{88--90} (\bibinfo{year}{2006}).

\bibitem{lazer2009computational}
\bibinfo{author}{Lazer, D.} \emph{et~al.}
\newblock \bibinfo{title}{Computational social science}.
\newblock \emph{\bibinfo{journal}{Science}} \textbf{\bibinfo{volume}{323}},
  \bibinfo{pages}{721} (\bibinfo{year}{2009}).

\bibitem{Kwak2010}
\bibinfo{author}{Kwak, H.}, \bibinfo{author}{Lee, C.}, \bibinfo{author}{Park,
  H.} \& \bibinfo{author}{Moon, S.}
\newblock \bibinfo{title}{{What is Twitter, a Social Network or a News Media?
  Categories and Subject Descriptors}}.
\newblock In \emph{\bibinfo{booktitle}{19th International Conference on the
  World Wide Web (WWW '10)}}, \bibinfo{pages}{591--600} (\bibinfo{year}{2010}).

\bibitem{Bakshy2015}
\bibinfo{author}{Bakshy, E.}, \bibinfo{author}{Messing, S.} \&
  \bibinfo{author}{Adamic, L.~A.}
\newblock \bibinfo{title}{{Exposure to ideologically diverse news and opinion
  on Facebook}}.
\newblock \emph{\bibinfo{journal}{Science}} \textbf{\bibinfo{volume}{348}},
  \bibinfo{pages}{1130--1132} (\bibinfo{year}{2015}).

\bibitem{Garciae1701172}
\bibinfo{author}{Garcia, D.}
\newblock \bibinfo{title}{Leaking privacy and shadow profiles in online social
  networks}.
\newblock \emph{\bibinfo{journal}{Science Advances}}
  \textbf{\bibinfo{volume}{3}} (\bibinfo{year}{2017}).

\bibitem{shirky2011political}
\bibinfo{author}{Shirky, C.}
\newblock \bibinfo{title}{The political power of social media: Technology, the
  public sphere, and political change}.
\newblock \emph{\bibinfo{journal}{Foreign Affairs}}
  \textbf{\bibinfo{volume}{90}}, \bibinfo{pages}{28--41}
  (\bibinfo{year}{2011}).

\bibitem{lotan2011arab}
\bibinfo{author}{Lotan, G.} \emph{et~al.}
\newblock \bibinfo{title}{The revolutions were tweeted: Information flows
  during the 2011 {T}unisian and {E}gyptian revolutions}.
\newblock \emph{\bibinfo{journal}{Int. J. Comm}} \textbf{\bibinfo{volume}{5}},
  \bibinfo{pages}{31} (\bibinfo{year}{2011}).

\bibitem{delvicario2016}
\bibinfo{author}{Del~Vicario, M.} \emph{et~al.}
\newblock \bibinfo{title}{The spreading of misinformation online}.
\newblock \emph{\bibinfo{journal}{Proc.\ Natl.\ Acad.\ Sci.\ U.\ S.\ A.}}
  \textbf{\bibinfo{volume}{113}}, \bibinfo{pages}{554--559}
  (\bibinfo{year}{2016}).

\bibitem{Castellano2009}
\bibinfo{author}{Castellano, C.}, \bibinfo{author}{Fortunato, S.} \&
  \bibinfo{author}{Loreto, V.}
\newblock \bibinfo{title}{{Statistical physics of social dynamics}}.
\newblock \emph{\bibinfo{journal}{Reviews of Modern Physics}}
  \textbf{\bibinfo{volume}{81}}, \bibinfo{pages}{591--646}
  (\bibinfo{year}{2009}).

\bibitem{Kramer2014}
\bibinfo{author}{Kramer, A.~D.}, \bibinfo{author}{Guillory, J.~E.} \&
  \bibinfo{author}{Hancock, J.~T.}
\newblock \bibinfo{title}{{Experimental evidence of massive-scale emotional
  contagion through social networks}}.
\newblock \emph{\bibinfo{journal}{Proceedings of the National Academy of
  Sciences of the United States of America}} \textbf{\bibinfo{volume}{111}},
  \bibinfo{pages}{8788--9790} (\bibinfo{year}{2014}).

\bibitem{Monsted2017}
\bibinfo{author}{M{\o}nsted, B.}, \bibinfo{author}{Sapie{\.{z}}y{\'{n}}ski,
  P.}, \bibinfo{author}{Ferrara, E.} \& \bibinfo{author}{Lehmann, S.}
\newblock \bibinfo{title}{{Evidence of Complex Contagion of Information in
  Social Media: An Experiment Using Twitter Bots}}.
\newblock \emph{\bibinfo{journal}{PLoS ONE}} \textbf{\bibinfo{volume}{12}},
  \bibinfo{pages}{e0184148} (\bibinfo{year}{2017}).

\bibitem{Jurgens2017}
\bibinfo{author}{Jurgens, D.}, \bibinfo{author}{Tsvetkov, Y.} \&
  \bibinfo{author}{Jurafsky, D.}
\newblock \bibinfo{title}{{Writer profiling without the writer's text}}.
\newblock In \emph{\bibinfo{booktitle}{Social Informatics. SocInfo 2017.
  Lecture Notes in Computer Science}}, vol. \bibinfo{volume}{10540},
  \bibinfo{pages}{537--558} (\bibinfo{year}{2017}).

\bibitem{Garcia2018}
\bibinfo{author}{Garcia, D.}, \bibinfo{author}{Goel, M.},
  \bibinfo{author}{Agrawal, A.~K.} \& \bibinfo{author}{Kumaraguru, P.}
\newblock \bibinfo{title}{{Collective aspects of privacy in the Twitter social
  network}}.
\newblock \emph{\bibinfo{journal}{EPJ Data Science}}
  \textbf{\bibinfo{volume}{7}} (\bibinfo{year}{2018}).

\bibitem{gruhl2004information}
\bibinfo{author}{Gruhl, D.}, \bibinfo{author}{Guha, R.},
  \bibinfo{author}{Liben-Nowell, D.} \& \bibinfo{author}{Tomkins, A.}
\newblock \bibinfo{title}{Information diffusion through blogspace}.
\newblock In \emph{\bibinfo{booktitle}{{WWW}}}, \bibinfo{pages}{491--501}
  (\bibinfo{organization}{ACM}, \bibinfo{year}{2004}).

\bibitem{bakshy2012role}
\bibinfo{author}{Bakshy, E.}, \bibinfo{author}{Rosenn, I.},
  \bibinfo{author}{Marlow, C.} \& \bibinfo{author}{Adamic, L.}
\newblock \bibinfo{title}{The role of social networks in information
  diffusion}.
\newblock In \emph{\bibinfo{booktitle}{{WWW}}}, \bibinfo{pages}{519--528}
  (\bibinfo{organization}{ACM}, \bibinfo{year}{2012}).

\bibitem{aral2009distinguishing}
\bibinfo{author}{Aral, S.}, \bibinfo{author}{Muchnik, L.} \&
  \bibinfo{author}{Sundararajan, A.}
\newblock \bibinfo{title}{Distinguishing influence-based contagion from
  homophily-driven diffusion in dynamic networks}.
\newblock \emph{\bibinfo{journal}{Proc.\ Natl.\ Acad.\ Sci.\ U.\ S.\ A.}}
  \textbf{\bibinfo{volume}{106}}, \bibinfo{pages}{21544--21549}
  (\bibinfo{year}{2009}).

\bibitem{centola2010spread}
\bibinfo{author}{Centola, D.}
\newblock \bibinfo{title}{The spread of behavior in an online social network
  experiment}.
\newblock \emph{\bibinfo{journal}{Science}} \textbf{\bibinfo{volume}{329}},
  \bibinfo{pages}{1194--1197} (\bibinfo{year}{2010}).

\bibitem{aral2012identifying}
\bibinfo{author}{Aral, S.} \& \bibinfo{author}{Walker, D.}
\newblock \bibinfo{title}{Identifying influential and susceptible members of
  social networks}.
\newblock \emph{\bibinfo{journal}{Science}} \textbf{\bibinfo{volume}{337}},
  \bibinfo{pages}{337--341} (\bibinfo{year}{2012}).

\bibitem{ver2012information}
\bibinfo{author}{Ver~Steeg, G.} \& \bibinfo{author}{Galstyan, A.}
\newblock \bibinfo{title}{Information transfer in social media}.
\newblock In \emph{\bibinfo{booktitle}{{WWW}}}, \bibinfo{pages}{509--518}
  (\bibinfo{organization}{ACM}, \bibinfo{year}{2012}).

\bibitem{borge2016dynamics}
\bibinfo{author}{Borge-Holthoefer, J.} \emph{et~al.}
\newblock \bibinfo{title}{The dynamics of information-driven coordination
  phenomena: A transfer entropy analysis}.
\newblock \emph{\bibinfo{journal}{Science Advances}}
  \textbf{\bibinfo{volume}{2}} (\bibinfo{year}{2016}).

\bibitem{CoverThomas}
\bibinfo{author}{Cover, T.~M.} \& \bibinfo{author}{Thomas, J.~A.}
\newblock \emph{\bibinfo{title}{Elements of Information Theory}}
  (\bibinfo{publisher}{John Wiley \& Sons}, \bibinfo{year}{2012}).

\bibitem{shannon1951prediction}
\bibinfo{author}{Shannon, C.~E.}
\newblock \bibinfo{title}{Prediction and entropy of printed english}.
\newblock \emph{\bibinfo{journal}{Bell Syst. Tech. J}}
  \textbf{\bibinfo{volume}{30}}, \bibinfo{pages}{50--64}
  (\bibinfo{year}{1951}).

\bibitem{brown1992estimate}
\bibinfo{author}{Brown, P.~F.}, \bibinfo{author}{Pietra, V. J.~D.},
  \bibinfo{author}{Mercer, R.~L.}, \bibinfo{author}{Pietra, S. A.~D.} \&
  \bibinfo{author}{Lai, J.~C.}
\newblock \bibinfo{title}{An estimate of an upper bound for the entropy of
  english}.
\newblock \emph{\bibinfo{journal}{Comput. Ling.}}
  \textbf{\bibinfo{volume}{18}}, \bibinfo{pages}{31--40}
  (\bibinfo{year}{1992}).

\bibitem{schurmann1996entropy}
\bibinfo{author}{Sch{\"u}rmann, T.} \& \bibinfo{author}{Grassberger, P.}
\newblock \bibinfo{title}{Entropy estimation of symbol sequences}.
\newblock \emph{\bibinfo{journal}{Chaos}} \textbf{\bibinfo{volume}{6}},
  \bibinfo{pages}{414--427} (\bibinfo{year}{1996}).

\bibitem{kontoyiannis_nonparametric_1998}
\bibinfo{author}{Kontoyiannis, I.}, \bibinfo{author}{Algoet, P.},
  \bibinfo{author}{Suhov, Y.~M.} \& \bibinfo{author}{Wyner, A.}
\newblock \bibinfo{title}{Nonparametric entropy estimation for stationary
  processes and random fields, with applications to english text}.
\newblock \emph{\bibinfo{journal}{{IEEE} Trans. Inf. Theory}}
  \textbf{\bibinfo{volume}{44}}, \bibinfo{pages}{1319--1327}
  (\bibinfo{year}{1998}).

\bibitem{song2010limits}
\bibinfo{author}{Song, C.}, \bibinfo{author}{Qu, Z.}, \bibinfo{author}{Blumm,
  N.} \& \bibinfo{author}{Barab\'asi, A.-L.}
\newblock \bibinfo{title}{Limits of predictability in human mobility}.
\newblock \emph{\bibinfo{journal}{Science}} \textbf{\bibinfo{volume}{327}},
  \bibinfo{pages}{1018--1021} (\bibinfo{year}{2010}).

\bibitem{schreiber2000measuring}
\bibinfo{author}{Schreiber, T.}
\newblock \bibinfo{title}{Measuring information transfer}.
\newblock \emph{\bibinfo{journal}{Phys. Rev. Lett.}}
  \textbf{\bibinfo{volume}{85}}, \bibinfo{pages}{461} (\bibinfo{year}{2000}).

\bibitem{staniek2008symbolic}
\bibinfo{author}{Staniek, M.} \& \bibinfo{author}{Lehnertz, K.}
\newblock \bibinfo{title}{Symbolic transfer entropy}.
\newblock \emph{\bibinfo{journal}{Phys. Rev. Lett.}}
  \textbf{\bibinfo{volume}{100}}, \bibinfo{pages}{158101}
  (\bibinfo{year}{2008}).

\bibitem{dunbar1993coevolution}
\bibinfo{author}{Dunbar, R.~I.}
\newblock \bibinfo{title}{Coevolution of neocortical size, group size and
  language in humans}.
\newblock \emph{\bibinfo{journal}{Behav. Brain. Sci.}}
  \textbf{\bibinfo{volume}{16}}, \bibinfo{pages}{681--694}
  (\bibinfo{year}{1993}).


\bibitem{Reka2000}
\bibinfo{author}{Albert, R.}, \bibinfo{author}{Jeong, H.} \&
  \bibinfo{author}{Barabasi, A.-L.}
\newblock \bibinfo{title}{{Error and attack tolerance of complex networks}}.
\newblock \emph{\bibinfo{journal}{Nature}} \textbf{\bibinfo{volume}{406}},
  \bibinfo{pages}{378--382} (\bibinfo{year}{2000}).

\bibitem{wasserman1994social}
\bibinfo{author}{Wasserman, S.} \& \bibinfo{author}{Faust, K.}
\newblock \emph{\bibinfo{title}{Social network analysis: Methods and
  applications}} (\bibinfo{publisher}{Cambridge university press},
  \bibinfo{year}{1994}).

\bibitem{de2013unique}
\bibinfo{author}{De~Montjoye, Y.-A.}, \bibinfo{author}{Hidalgo, C.~A.},
  \bibinfo{author}{Verleysen, M.} \& \bibinfo{author}{Blondel, V.~D.}
\newblock \bibinfo{title}{Unique in the crowd: The privacy bounds of human
  mobility}.
\newblock \emph{\bibinfo{journal}{Sci. Rep.}} \textbf{\bibinfo{volume}{3}},
  \bibinfo{pages}{1376} (\bibinfo{year}{2013}).

\bibitem{de2015unique}
\bibinfo{author}{de~Montjoye, Y.-A.}, \bibinfo{author}{Radaelli, L.},
  \bibinfo{author}{Singh, V.~K.} \& \bibinfo{author}{Pentland, A.}
\newblock \bibinfo{title}{Unique in the shopping mall: On the reidentifiability
  of credit card metadata}.
\newblock \emph{\bibinfo{journal}{Science}} \textbf{\bibinfo{volume}{347}},
  \bibinfo{pages}{536--539} (\bibinfo{year}{2015}).

\bibitem{pariser2011filter}
\bibinfo{author}{Pariser, E.}
\newblock \emph{\bibinfo{title}{The filter bubble: What the Internet is hiding
  from you}} (\bibinfo{publisher}{Penguin UK}, \bibinfo{year}{2011}).

\bibitem{mosteller1963inference}
\bibinfo{author}{Mosteller, F.} \& \bibinfo{author}{Wallace, D.~L.}
\newblock \bibinfo{title}{Inference in an authorship problem: A comparative
  study of discrimination methods applied to the authorship of the disputed
  federalist papers}.
\newblock \emph{\bibinfo{journal}{Journal of the American Statistical
  Association}} \textbf{\bibinfo{volume}{58}}, \bibinfo{pages}{275--309}
  (\bibinfo{year}{1963}).

\bibitem{katz1987estimation}
\bibinfo{author}{Katz, S.}
\newblock \bibinfo{title}{Estimation of probabilities from sparse data for the
  language model component of a speech recognizer}.
\newblock \emph{\bibinfo{journal}{IEEE transactions on acoustics, speech, and
  signal processing}} \textbf{\bibinfo{volume}{35}}, \bibinfo{pages}{400--401}
  (\bibinfo{year}{1987}).

\bibitem{bengio2003neural}
\bibinfo{author}{Bengio, Y.}, \bibinfo{author}{Ducharme, R.},
  \bibinfo{author}{Vincent, P.} \& \bibinfo{author}{Jauvin, C.}
\newblock \bibinfo{title}{A neural probabilistic language model}.
\newblock \emph{\bibinfo{journal}{Journal of machine learning research}}
  \textbf{\bibinfo{volume}{3}}, \bibinfo{pages}{1137--1155}
  (\bibinfo{year}{2003}).

\bibitem{shalizi2011homophily}
\bibinfo{author}{Shalizi, C.~R.} \& \bibinfo{author}{Thomas, A.~C.}
\newblock \bibinfo{title}{Homophily and contagion are generically confounded in
  observational social network studies}.
\newblock \emph{\bibinfo{journal}{Sociological methods \& research}}
  \textbf{\bibinfo{volume}{40}}, \bibinfo{pages}{211--239}
  (\bibinfo{year}{2011}).

\bibitem{granger1969investigating}
\bibinfo{author}{Granger, C. W.~J.}
\newblock \bibinfo{title}{Investigating causal relations by econometric models
  and cross-spectral methods}.
\newblock \emph{\bibinfo{journal}{Econometrica}} \textbf{\bibinfo{volume}{37}},
  \bibinfo{pages}{424--438} (\bibinfo{year}{1969}).

\bibitem{TwitterAPI}
\bibinfo{title}{{Twitter REST APIs}}.
\newblock \bibinfo{howpublished}{Available from:
  \url{https://dev.twitter.com/rest/public}} (\bibinfo{year}{2016}).
\newblock \bibinfo{note}{Accessed: 2016-07-07}.

\bibitem{BotometerApi}
\bibinfo{title}{{Botometer API}}.
\newblock \bibinfo{howpublished}{Available from:
  \url{https://botometer.iuni.iu.edu/}} (\bibinfo{year}{2016}).
\newblock \bibinfo{note}{Accessed: 2016-07-07}.

\bibitem{botornot1}
\bibinfo{author}{Varol, O.}, \bibinfo{author}{Ferrara, E.},
  \bibinfo{author}{Davis, C.~A.}, \bibinfo{author}{Menczer, F.} \&
  \bibinfo{author}{Flammini, A.}
\newblock \bibinfo{title}{Online human-bot interactions: Detection, estimation,
  and characterization}.
\newblock In \emph{\bibinfo{booktitle}{ICWSM}} (\bibinfo{year}{2017}).

\bibitem{botornot2}
\bibinfo{author}{Davis, C.~A.}, \bibinfo{author}{Varol, O.},
  \bibinfo{author}{Ferrara, E.}, \bibinfo{author}{Flammini, A.} \&
  \bibinfo{author}{Menczer, F.}
\newblock \bibinfo{title}{{BotOrNot}: A system to evaluate social bots}.
\newblock In \emph{\bibinfo{booktitle}{WWW Developers Day}}
  (\bibinfo{year}{2016}).

\bibitem{botornot3}
\bibinfo{author}{Ferrara, E.}, \bibinfo{author}{Varol, O.},
  \bibinfo{author}{Davis, C.~A.}, \bibinfo{author}{Menczer, F.} \&
  \bibinfo{author}{Flammini, A.}
\newblock \bibinfo{title}{The rise of social bots}.
\newblock \emph{\bibinfo{journal}{Communications of the ACM}}
  \textbf{\bibinfo{volume}{59}} (\bibinfo{year}{2016}).

\bibitem{botornot4}
\bibinfo{author}{Subrahmanian, V.~S.} \emph{et~al.}
\newblock \bibinfo{title}{The {DARPA} {T}witter {B}ot {C}hallenge}.
\newblock \emph{\bibinfo{journal}{Computer}} \textbf{\bibinfo{volume}{49}},
  \bibinfo{pages}{38--46} (\bibinfo{year}{2016}).

\bibitem{243444}
\bibinfo{author}{Ziv, J.} \& \bibinfo{author}{Merhav, N.}
\newblock \bibinfo{title}{A measure of relative entropy between individual
  sequences with application to universal classification}.
\newblock \emph{\bibinfo{journal}{{IEEE} Trans. Inf. Theory}}
  \textbf{\bibinfo{volume}{39}}, \bibinfo{pages}{1270--1279}
  (\bibinfo{year}{1993}).

\end{thebibliography}
\end{document}